\documentclass[useAMS,usenatbib]{mn2e}

\usepackage{amsmath,amssymb,hyperref,epstopdf,graphicx,aas_macros}
\usepackage{sidecap}
\usepackage{caption}
\usepackage[figuresright]{rotating}

\newcommand{\fflex}{{\mathcal F}}
\newcommand{\gflex}{{\mathcal G}}

\usepackage{lipsum}

\title[Cluster substructure with gravitational flexion]{Reconstruction of small-scale galaxy cluster substructure with lensing flexion}
\author[Cain et al.]{Benjamin Cain$^1$\thanks{E-mail: \texttt{bmcain@ucdavis.edu}}, Maru\v{s}a Brada\v{c}$^1$, and Rebecca Levinson$^2$ \\
$^1$University of California Davis, Department of Physics, One Shields Ave., Davis, CA 95616 \\
$^2$Massachusetts Institute of Technology, Department of Physics, 77 Massachusetts Ave., Cambridge, MA 02139}

\begin{document}
\date{\today}
\maketitle

\begin{abstract}
We present a reconstructions of galaxy-cluster-scale mass distributions from simulated gravitational lensing data sets including strong lensing, weak lensing shear, and measurements of quadratic image distortions -- flexion.  The lensing data is constructed to make a direct comparison between mass reconstructions with and without flexion.  We show that in the absence of flexion measurements, significant galaxy-group scale substructure can remain undetected in the reconstructed mass profiles, and that the resulting profiles underestimate the aperture mass in the substructure regions by $\sim25-40\%$.  When flexion is included, subhaloes down to a mass of $\sim3\times10^{12}$ M$_\odot$ can be detected at an angular resolution smaller than 10\arcsec.  Aperture masses from profiles reconstructed with flexion match the input distribution values to within an error of $\sim13\%$, including both statistical error and scatter.  This demonstrates the important constraint that flexion measurements place on substructure in galaxy clusters and its utility for producing high-fidelity mass reconstructions.
\end{abstract}


%%%%%%%%%%%%%%%%%%%%%%
%%%%%%%%%%%%%%%%%%%%%%
\section{Introduction}
The distribution of mass structure in the Universe provides a fundamental test for any plausible cosmological model.  Empirical data from multiple experiments have robustly constrained cosmology to be consistent with a flat $\Lambda$CDM model \citep{2013ApJS..208...19H,2008ApJ...686..749K,2013arXiv1303.5076P,2005ApJ...633..560E}.  But to further understand the constituents of this cosmology, in particular the dark matter, we must investigate the properties of the structures which form from the primordial mass distribution at many scales.  To understand the interaction between dark matter and baryonic matter, or to understand the self-interactions between dark matter particles we must measure the distribution of structures on relatively small scales, meaning measuring structures down to galaxy scales where the properties of both baryons and dark matter influence the evolution of structure.  As interactions between dark matter and baryons are more important to the structure of haloes for more non-linear overdensities, measurements of substructures inside of larger halos (e.g., substructures within galaxy clusters) provide an important window into the nature of dark matter.

For a $\Lambda$CDM cosmology, the particular initial conditions (e.g., the primordial matter power spectrum) as well as the particular properties of the dark matter -- both its self-interactions and its interaction with baryons, if any -- will affect the structures that are formed.  This includes the mass function of virialized  dark matter haloes \citep{2008ApJ...688..709T} and the spatial and mass distributions of subhaloes \citep{2005ApJ...618..557N,2004ApJ...609...35K}.  The existence of roughly self-similar haloes at all scales is a robust prediction of $\Lambda$CDM \citep{1999ApJ...524L..19M}, making the detection of galaxy cluster substructures an important cosmological test.  The fraction of a cluster's mass which is associated with substructures as well as the other subhalo properties (e.g., density profiles, ellipticities, etc.) can be compared to subhaloes produced in cosmological $N$-body simulations to further constrain dark matter properties. Recent cosmological simulations of clusters with self-interacting  dark matter (SIDM) have found that dark matter interactions affect both the shape of haloes, with more spherical profiles and flatter halo cores as a result of a higher interaction cross-section for galaxy-scale haloes, as well as reducing the number of subhaloes of a given mass \citep{2006ApJ...646..815S,2013MNRAS.430...81R,2013MNRAS.430..105P}.

To measure the mass distribution of substructures within galaxy clusters we cannot use statistical measures of cluster properties, such as empirical scaling relations \citep[see, e.g.,][for a review of these relations]{2013SSRv..177..247G}.  These relationships are useful for large samples of clusters but the intrinsic scatter between individual clusters limits their predictive power. Neither can we use baryonic tracers of structure due to uncertainty in the interactions between baryons and dark matter.  However, gravitational lensing, and particularly  combining strong lensing deflections and linear image distortions (weak lensing shear and convergence) has been successfully shown to produce detailed mass reconstructions of galaxy clusters and resolved complicated, asymmetric morphologies \citep{2006ApJ...648L.109C,2007MNRAS.376..180N,2009ApJ...693..970N}.  

For typical cluster imaging datasets, the correlation of spatially-separated galaxy shapes with uncertain redshifts produces an effective smoothing length of $\gtrsim10\arcsec$ away from the central region of the cluster which is constrained by strong lensing.  Simulations indicate that at least half of dark matter subhaloes will exist outside of the Einstein radius of a the main halo \citep{2005ApJ...618..557N}.  In this large portion of the cluster, at least 99\% of the projected area within the cluster virial radius, substructures cannot be efficiently detected in galaxy clusters without additional information or the investment of prohibitively large amounts of telescope time.

In order to overcome these limitations, to improve the resolution of reconstructed mass distributions, and thus to probe small-scale mass structure, we includie higher-order lensing effects.  In regions where the reduced lensing shear $g=\gamma/(1-\kappa)$ is a significant fraction of unity, as is the case within a radius of $\sim$60\arcsec\ from the cluster center for a typical galaxy cluster, nonlinear image distortions become significant and the linear distortion approximation becomes increasingly less valid as the shear increases \citep[see][for a review of linear weak lensing]{2001PhR...340..291B}, which also degrades the reliability of galaxy ellipticity as an estimator for shear with standard approaches \citep{2010A&A...510A..75M}.  Quadratic image distortions due to non-negligible third derivatives of the lensing potential, called flexion, can be observed and quantitatively measured in this regime.  Flexion distortions cause intrinsically elliptical galaxy images to be bent by into ``banana-shaped'' arclets, augmenting the image distortions due to linear shear and magnification.  The two lensing fields which cause flexion distortions are sourced by the gradients of the convergence and shear fields.  This has the practical effect of making flexion a sensitive probe of small-scale variations in the surface mass density.  Limits on the measurable flexion also constrain the amount of possible substructure in a region where no flexion is significantly detected.

In this paper we present a method for reconstructing a lensing mass distribution using multiple image positions (strong lensing), ellipticity measurements (weak lensing shear), and flexion measurements. This significantly improves the sensitivity of the reconstruction to small mass substructures; we are able to resolve structures to a significantly finer spatial resolution and a lower mass threshold than with strong lensing and shear alone.  We are able to directly detect structures with mass $\sim 9\times10^{12}$M$_\odot$ which are not detected without the flexion measurements, even in the case where they are located far from the strong lensing regime.

Throughout we employ a standard flat cosmological model with $\Omega_m=0.3$, $\Omega_\Lambda=0.7$, and $H_0=70$ km/s/Mpc.  We also employ standard complex notation for lensing fields, with $\nabla=\partial_1+i\partial_2$ being the derivative operator acting in the image plane.

\section{Mass Reconstruction Method}\label{sec:reconmethod}
Our mass reconstruction method is an expansion of the method described in \citet{Bradac:2009bk}, which itself was an improvement of previous methods \citep{Bradac:2005ex,2006ApJ...652..937B}.  We review the basic concept of the method here to place our expansion in context.  We include five complex lensing fields, all of which yield measurable lensing distortions which are inputs into our reconstruction.  All five fields arise as linear combinations of the first through third derivatives of a single, real-valued lensing potential $\psi$ sourced by the lens mass distribution.  These lensing fields are the deflection $\alpha=\nabla\psi$, the convergence $\kappa=\frac{1}{2}\nabla\nabla^*\psi$, the shear $\gamma=\frac{1}{2}\nabla^2\psi$, 1-flexion (comatic flexion) $\fflex=\nabla\kappa$, and 3-flexion (trefoil flexion) $\gflex=\nabla\gamma$.  Here we have used complex notation for both the fields and the derivative operators, and the star operator denotes the complex conjugate.  The positions of the multiple images in a strong lensing system constrain the deflection, and the ellipticity of lensed images of background galaxies constrains the shear and the convergence \citep[see][for a review of strong lensing and shear constraints on lens mass distributions]{2001PhR...340..291B}.

Both 1-flexion and 3-flexion are constrained by measurements of non-linear distortions due to gradients in the convergence and shear.  Since the convergence is the surface mass density of the lens scaled to the critical lensing density, measurements of image distortions due to flexion provide an important direct probe of mass gradients, and therefore small-scale mass structures.  The flexion fields can be measured using several methods available in the literature \citep{2005ApJ...619..741G,2007ApJ...660.1003G,Okura:2009cw,Schneider:2008ho}. In our mass reconstruction we adopt the flexion formalism as defined by \citet[hereafter CSB]{2011ApJ...736...43C}.  Our reconstruction method assumes measurements of the reduced flexion fields $\Psi_1=\fflex/(1-\kappa)$ and $\Psi_3=\gflex/(1-\kappa)$ from lensed source images using the Analytic Image Model (AIM) approach, rather than calculating flexion estimates from image characteristics like moments or shapelet weights.

Using the estimators for each of the lensing fields obtained from analyzing cluster data images, we constrain a model lensing potential and the mass distribution which corresponds to that potential.  Our model is defined as lens potential values $\psi_k$ on a ``mesh'' of points $\theta_k$ to be constrained by the data.  The value of the derivatives of the model potential at an arbitrary point $\theta_j$ are calculated using a generalized finite differencing method using appropriate linear combination of nearby potential values on the mesh to give the desired derivative.  The coefficients of this linear transformation are determined using a singular-value decomposition (SVD) as in \citet{Bradac:2009bk} to yield a system of equations relating the lensing potential values in the neighborhood of an arbitrary point in the field to the derivatives of the lensing potential:
\begin{equation}
	\partial_1^{n}\partial_2^{m}\psi(\theta_j)=\sum_{k}a_{jk}(n,m)\psi_k,
\end{equation}
where $n$ and $m$ are the derivative orders for each dimension, $k$ indexes the mesh points and $a_{jk}(n,m)$ is the SVD matrix coefficient for the weight of the potential value contribution of the $k^{\textrm{th}}$ mesh point to the $n,m$ derivative of the potential at $\theta_j$. As the value for each lensing field is a linear combination of potential field derivatives, this scheme calculates each lensing field at any point in the field as a linear combination of model potential values from neighboring mesh points to $\theta_j$.  To constrain the model at the location of a lensed image, the model value for the lensing field(s) which have estimators derived from the image can be calculated and compared to the model.

We then determine the values of $\psi_k$ which best fit the data by minimizing a model penalty function ($\chi^2$) which includes contributions from weak lensing shear, strong lensing, and flexion to penalize models which do not produce the measured lensing distortions.  The form of the strong and weak lensing penalty functions are identical to those of \citet{Bradac:2009bk}.  The value of the flexion penalty function is determined by the reduced flexion fields $\Psi_1^i$ and $\Psi_3^i$ (where $i$ indexes the $N_F$ galaxies with measured flexion fields), and the model values for the flexion and convergence fields $\fflex^i$, $\gflex^i$, and $\kappa^i$ calculated for the location of that galaxy:
\begin{equation}
	\chi^2_{FL}=\sum_{i}^{N_F}\left[\frac{1}{\sigma_{\fflex,i}^2}\left|\Psi_1^i-\frac{\fflex^i}{1-\kappa^i}\right|^2 + \frac{1}{\sigma_{\gflex,i}^2}\left|\Psi_3^i-\frac{\gflex^i}{1-\kappa^i}\right|^2\right].
\end{equation}
Here, $N_{F}$ is the number of galaxy images with flexion measurements.  As described above, the model flexion and convergence values are determined from a linear combination of mesh potential values.  The uncertainties $\sigma_{\fflex,i}$ and $\sigma_{\gflex,i}$ include both intrinsic flexion scatter and measurement error.  This is the same basic form for the penalty function as described in \citet{Er:2010uo}, but we explicitly formulate it in terms of the reduced flexion estimators $\Psi_1$ and $\Psi_3$ from CSB.  

As in \citet{Bradac:2009bk}, we include in our full penalty function regularization terms, which prevent the highly-nonlinear nature of the penalty function's dependence on the mesh point potential values from driving the model to assume values wildly discrepant from the initial conditions during the reconstruction.  This regularization is particularly important for mesh points in regions that are weakly constrained by the data and therefore more susceptible to numerical effects in the reconstruction.  We choose to regularize only on the values of convergence and shear, meaning that we penalize model potential values at each mesh point which require large pixel-to-pixel variations in the lensing fields fields relative to a reference value.  Note that this does not mean that the model cannot vary from its initial conditions!  Instead, we are requiring that the deviation from the initial conditions be driven by the data, rather than numerical effects.  We iterate the reconstruction process and reset the reference value to the output of the previous iteration.  This causes the model to step smoothly away from the initial guess in a controlled and converging way.  Without regularization, the highly nonlinear dependence of the reduced lensing fields prevents the reconstruction process from finding a solution which minimizes the penalty function.

Our regularization penalty function for convergence is defined over as a sum over the $N_{mesh}$ mesh points located at $\theta_k$ as
\begin{equation}\label{eq:regchisq}
	\chi^2_{reg,\kappa}=\sum_{k}^{N_{mesh}}\eta^{(\kappa)}\left(\kappa(\theta_k)-\kappa^{(0)}(\theta_k)\right)^2,
\end{equation}
with similar terms for each component of the shear field. The value of the regularization weights $\eta^{(\kappa)}$, $\eta^{(\gamma_{1})}$, and $\eta^{(\gamma_{2})}$, are determined empirically.  We test a range of values for each reconstruction to not only find an appropriate value which prevents spurious nonlinear numerical artifacts while still allowing the reconstruction to deviate from initial conditions as indicated by data, but also to ensure that any reconstruction is stable with respect to variations the regularization coefficients.  We also select values which place the total regularization penalty per mesh point (as a proxy for the regularization degrees of freedom) to be approximately equal that of the penalty per degree of freedom for each data portion of the full penalty function (i.e., strong, weak, or flexion lensing).

\section{Testing the Reconstruction Method}
With flexion included into the mass reconstruction, we test our method using simulated measurement catalogs.  We generate realistic lensing estimator values, including realistic measurement errors, from an underlying mass model which includes substructure.  We then reconstruct the mass distribution using all the available data, and compare that reconstruction to a modified reconstruction which incorporates the same data for strong and weak lensing shear, but omits the flexion data.  This allows us to make direct comparisons between reconstructions and evaluate the contribution that flexion makes to the fidelity of the final mass reconstructions.  In particular, we are interested in finding substructure in the mass distribution which would be unobserved without the inclusion of flexion data.

This latter point is important for how we construct our input simulated lensing measurements.  It has already been well established that strong lensing is a sensitive probe of substructures.  The goal of the mass reconstructions using our simulated measurement catalogs is to test whether realistic substructures in cluster lenses can exist which are detectable using flexion but not without.  As we will show, this is indeed the case.

\subsection{Input Lensing Mass Distribution}
Our input mass distribution is created to mimic a galaxy-cluster sized halo with a small number of substructures, each simulated using a nonsingular isothermal ellipse model.  We set the scale of the large central halo to be comparable with a galaxy cluster velocity dispersion of approximately 1500 km/s.  This is a large cluster, similar to those in the CLASH \citep{2012ApJS..199...25P} and Hubble Frontier Fields initiative\footnote{ http://www.stsci.edu/hst/campaigns/frontier-fields} (PI Lotz: Program ID 13495) samples.  We also include two substructures in the lens within a radius of 1.5 arcminutes of the cluster center.  These subhaloes are also nonsingular isothermal ellipses with varying masses from group scale down to large-galaxy scale (see Appendix A for tabulated subhalo masses).  We construct six different data sets corresponding to substructured cluster haloes.

The positioning of the halo mass substructures was not done in a fine-tuned way, i.e., we did not select the location or masses of the substructures in any way to enhance the effect of the substructures on the resultant lensing morphology.  Instead, we use a conservative approach and restrict the possible locations such that the strong lensing morphology does not initially indicate that there is significant substructure in the lens.  The intent of this work is to characterize the differential effect that flexion makes on the resulting mass distribution.  Thus, we restrict the substructure halo masses and positions to be outside the approximate radius of the critical curve so that the additional substructure is not easily detected using the strong lensing data alone.  Substructures which are too massive or near the critical curves create significant deviations in the location, morphology, and magnification of strong lensing images.  This can be an extremely sensitive probe of substructure \citep[e.g.][]{2012Natur.481..341V}, however our goal is to make a total census of the substructure in galaxy clusters.

\subsection{Simulated Lensing Estimators}
We simulate lensing measurement catalogs to test the mass reconstruction method and the effect of including flexion.  Using the main halo and smaller substructures we define as above, we analytically calculate the appropriate lensing field estimators from the total mass profile using the full lens equation and its derivatives.  We assume a single, moderate redshift of $z_{lens}=0.3$ for the lens halo (including its substructures) in all of the simulated datasets. Independent random noise is added to each simulated measurement with scatter chosen to match observational constraints from single orbit Hubble Space Telescope Advanced Camera for Surveys (\emph{HST}/ACS) cluster observations.  For the simulations presented here, we employ an input lensing mass distribution consisting of three non-singular isothermal ellipse (NIE) halo models for computational efficiency in constructing the simulated measurements and a simple comparison model.

To determine our simulated measurement values based on the lensing mass distribution, the source-lens geometry must be determined along the line of sight in addition to the plane of the sky.  This is because the lensing fields are all dependent on the redshift of the lensing mass as well as the redshift of the source object.  We randomly select the redshift of each lensed source from a gamma distribution
\begin{equation}
	p(z)=\frac{z^2}{2z_0^3}\exp(-z/z_0)
\end{equation}
with $z_0=2/3$.
%, yielding a mean redshift of 2
  This is the same distribution as used in previous work to emulate the true distribution of sources \citep{1996ApJ...466..623B,2007A&A...468..823S,2004A&A...424...13B}.  We draw a new random redshift for each flexion and weak lensing source image, and each strong lensing image system has its own redshift as well.  Redshift errors on the weak lensing/flexion sources are added with $\sigma_z=0.05(1+z)$.  The strong lensing systems do not have redshift errors, to reflect the spectroscopic redshifts which are available for many multiply imaged galaxies.   We describe the measurement generation procedure for each type of catalog (flexion, weak lensing, and strong lensing) below.

To generate our flexion measurement catalog we select a random distribution of source positions for the flexion and leak lensing shear measurements.  The positions are selected uniformly across a square field 3\farcm5 on a side to a density of 80 galaxies/arcmin$^2$.  For the flexion measurements we calculate the reduced flexion fields $\Psi_1$ and $\Psi_3$ as defined in CSB at the position of each of the sources.  We add Gaussian noise to each component of the flexion measurements with a standard deviation of 0.03 arcsec$^{-1}$.  These noise levels are motivated by previous studies of intrinsic flexion \citep[CSB]{2007ApJ...660.1003G}.  

We then remove from the flexion dataset any sources where the magnitude of the measured flexion exceeds 1.0 arcsec$^{-1}$.  This ceiling on the measured flexion reflects how the quadratic expansion of the lens equation which defines the flexion fields is only valid when the flexion correction is small relative to the shear distortion.  In a real dataset, images with flexion values in these regimes would either be identifiable as strongly-lensed arcs and thus in the strong lensing dataset instead of the flexion catalog, or would be removed during image fitting as in CSB.  There the authors found that for flexion values this large the AIM image fitting did not produce accurate results, but the properties of the AIM covariance matrices allowed for efficient removal of these catastrophic outlier images.  Our limit on the flexion values is a uniform approximation of the limit that would occur in real data: the directly observable distortion in a lensed image due to flexion is quantified by the unitless product of the image scale (such as the half-light radius) and the reduced flexion field, rather than the more physically relevant reduced flexion field alone.  Because flexion is a quadratic effect, measuring flexion requires that the lensed galaxy image be larger than is required for ellipticity - in practice, this means 2-3 times the point-spread function size along the long axis of the image. In order to fully include varying image size and its effect on flexion estimation and the resulting mass reconstruction, a better understanding of intrinsic flexion and its correlation to galaxy size and magnitude must be achieved.  However, the simpler cut on the value of the flexion field is a sufficient approximation for our purpose.

The weak lensing shear and strong lensing datasets are constructed from the reduced shear and deflections fields.  The source locations for the shear are randomly distributed through the simulated field and locations with reduced shears nearing unity are removed (for the same reason as large flexion values are removed).  The selection of the source position for the strong lensing systems is not completely random, however.  For each system we initially draw a random position from a small field in the source plane centered on the main halo core and solve the lens equation to find the image positions.  We then perform a visual inspection and remove systems if they show obvious signs of substructure.  With some source positions relative to the distribution of mass in the lens, multiple image systems will form around the smaller substructure.  We reject these systems, as the utility of strong lensing for substructure detection is already well established.  Instead, we select only image systems that are representative of a single, central halo and do not have obvious substructure included.

\subsection{Initial Conditions and Regularization}
Our approach to reconstructing the mass distribution from the simulated measurements is the same approach we would take using real data.  We determine initial conditions for our model fitting, set a mesh of points where the model potential will be defined, and reconstruct the lensing potential all using data-driven conditions.  We perform several reconstructions with different values for some of the less-strongly-constrained input parameters (e.g., the regularization weights).   We evaluate the output reconstructions, and in particular we evaluate the behavior of the penalty function values as the reconstruction converges, to determine the best values for these parameters. We do not see a strong dependence on the particular values - small changes do not significantly affect the output reconstruction.

The first step in our reconstruction is to assign initial conditions for the reconstruction which are motivated by the data.  We use an analytic model to set the initial model lensing potential values which, in turn, determine the initial value of the penalty function and its gradient.  As was shown previously \citep{2004A&A...424...13B,2005A&A...437...39B}, the final reconstruction using this method is not strongly dependent the particular initial model used provided that the regularization is chosen appropriately.  We employ a simple, parametric, axisymmetric model as our initial guess.  We take a single, nonsingular isothermal sphere as our initial model, locating the center of the NIS at the average position of the images in the strong lensing data set.  The mean of the distances from these strong lensing images from their average position defines an approximate Einstein radius for the overall lensing system.  Using the approximated Einstein radius estimate we construct our initial model from a NIS profile.  These simple, reasonable approximations provide data-driven initial conditions.

Another important input parameter to the mass reconstruction is the weight of each component of the regularization in the penalty function ($\eta_\kappa$, etc., from Eq.~\ref{eq:regchisq}), which defines the regularization weight relative to the strong lensing, weak lensing shear, and flexion contributions.  As noted in \S\ref{sec:reconmethod}, too little weight given to regularization (small $\eta$ values) yields reconstructions with unrealistically large amounts of small-scale spatial variation as well as aphysical lensing potential values, such as negative convergence.  These numerical artifacts of the nonlinear lens inversion prevent the optimization algorithm from finding the appropriate minimum as it iterates.  Excessive weight to the regularization (large $\eta$) causes the reconstruction to simply return a model which is minimally different from the initial mass model, preventing the data from informing the reconstructed mass distribution.  Because it is not possible to determine the most appropriate $\eta$ values \emph{a priori}, and because the most appropriate specific value for the regularization weight will depend not only on the details of the data available but also on the particular mesh geometry chosen, we determine $\eta$ empirically by reconstructing with a range of values.  By comparing the $\chi^2$ values for each part of the penalty function, we are able to separate successful reconstructions from those which are over- or under-regularized.

By evaluating the $\chi^2$ values output by these reconstructions we determine the approximate best combination of values for $\sigma_{NIS}$ and regularization weight to use for the given dataset.  To compare different reconstructions, we consider both the total $\chi^2$ value, as an indicator of the overall fitness of the reconstruction, and the individual contributions of the different datasets.  Reconstructions which yield values of $\chi^2/$d.o.f.$\approx1$ for each of the three datasets are preferred, even in cases where that increases both the overall and regularization $\chi^2$ values.  Furthermore, we prefer models which reproduce the strong lensing data more precisely above those which minimize the weak lensing or flexion penalty functions, provided that there are no obvious signs of significant pixel to pixel variations.  This is because strong lensing data have very small measurement errors, in fact smaller than the error we use in the strong lensing penalty function which we inflate in order to more easily locate the penalty function minimum. Thus if the strong lensing image systems are robustly identified in a given cluster, we can reasonably require that the solution have a strong lensing penalty value of effectively zero.  The reconstruction which meets these criteria while minimizing the overall $\chi^2$ value is the reconstruction which we select as our solution.

In the presence of a significantly substructured mass distribution, the reconstruction will not converge with a single halo initial model, and will plateau to $\chi^2$ values of $\sim$5-10 (or more) per degree of freedom after several iterations.  This is an indication of an initial guess which is far from the solution, and the highly non-linear nature of the reconstruction process requires and update to the initial conditions based off this information.  We update the initial conditions iteratively, occasionally adding additional haloes at the location of and sizes of substructures detected in the non-converged reconstruction in order to arrive at a good fit.  See \S\ref{results} for a discussion of our substructure detection method.

\subsubsection{Reconstruction Model Mesh}
We distribute mesh points over the ``observed field'', which in this case is the full area around the mean of the strong lensing image locations in a square region 3\farcm5 on a side (mimicking the \emph{HST/ACS} field of view).  These mesh points are where the model lensing potential and its derived lensing fields are defined.  The ``base'' mesh is defined by a rectangular, regular grid of points evenly defined across our field.  We refine this grid in  a circular region centered on the field center-point.  The radius of this circular region is 1.5 times the mean distance of the strong lensing images from the average strong lensing image location.  Within this radius, we distribute mesh points in a rectilinear grid at twice the base density.

We further refine in regions within a 3\arcsec\ radius of a strong lensing image position with four times the base density.  This refinement is necessary to accurately constrain the source positions for the strong lensing systems.  Numerical error in the calculation of the gradient of the model lensing potential due to a paucity of mesh points near strong lensing images can dominate the strong lensing contribution to the penalty function, which will preclude convergence and accurate reconstruction.

This nested mesh structure places increasing numbers of mesh points where the mass distribution is expected to be varying more rapidly, as well as where we have tighter constraints from the lensing data, and therefore a higher density of mesh points for the finite difference derivatives used to calculate the model lensing field values enhances the accuracy of the reconstruction output.

In the reconstructions presented here, we begin with a base mesh density of roughly 70 mesh points/arcmin$^2$, corresponding to about one galaxy image per mesh point, with the density of mesh points higher in the refinement regions.  The optimal configuration will have increased mesh point density wherever the potential is changing rapidly, and will not have any changes in refinement level of more than a single step.

\section{Results}\label{results}
Using the methods described above, we reconstruct the mass distribution for each of the simulated cluster datasets, and bootstrap the simulation 1 catalogs and reconstruct each of those.  Here we present our results, first in a detailed investigation of simulation 1 and the statistics of the reconstructions after bootstrap-resampling of the lensing catalogs, then more broadly for the rest of the cluster data sets.  In discussing these results, we will refer to locations within the reconstructed field by celestial directions (north being up and east being to the left), as we would were these actual cluster fields and not arbitrarily oriented simulations.

\subsection{Simulation 1}
\begin{figure}
	\begin{tabular}{c}
		\includegraphics[width=0.45\textwidth]{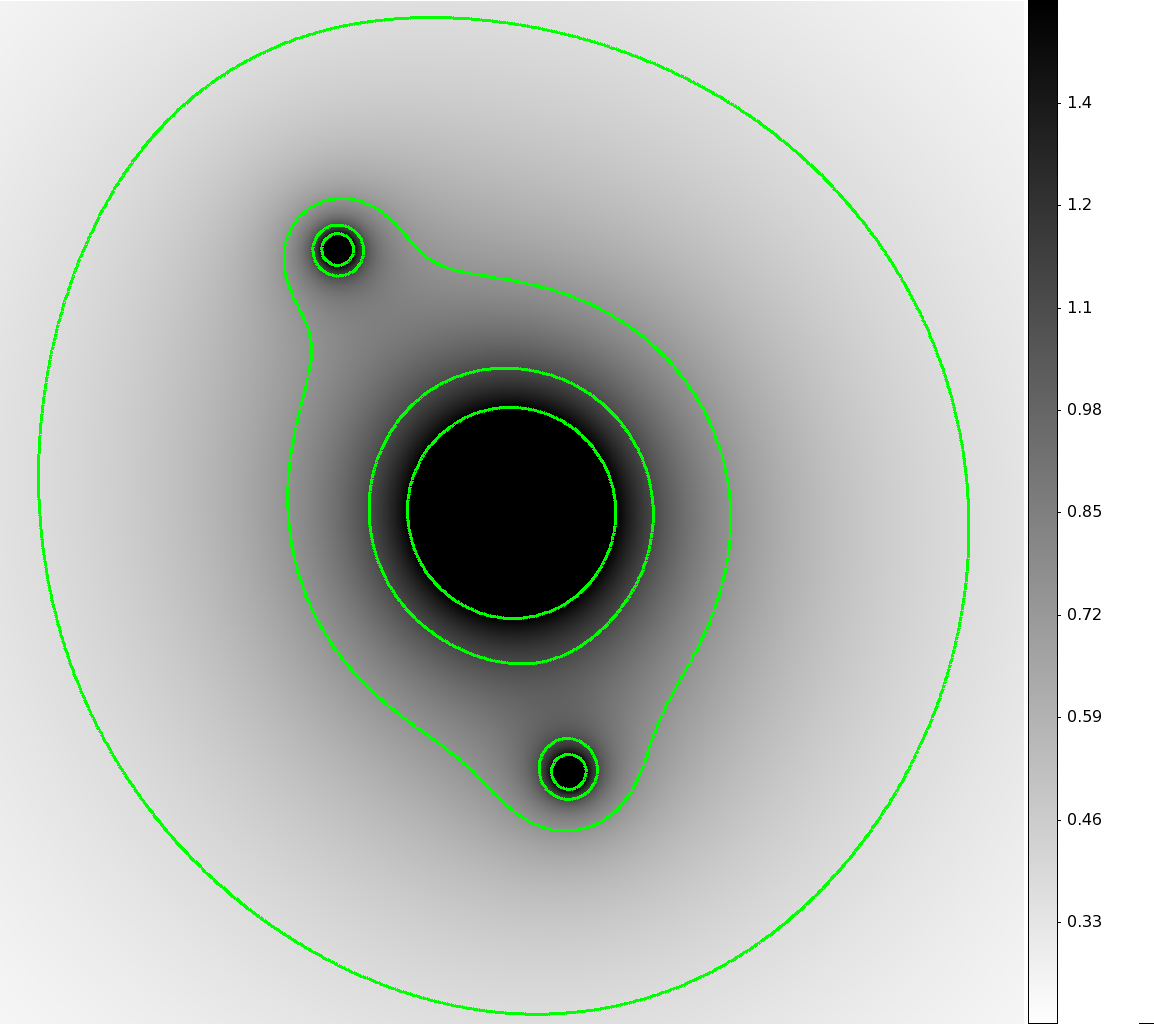} \\
		\includegraphics[width=0.45\textwidth]{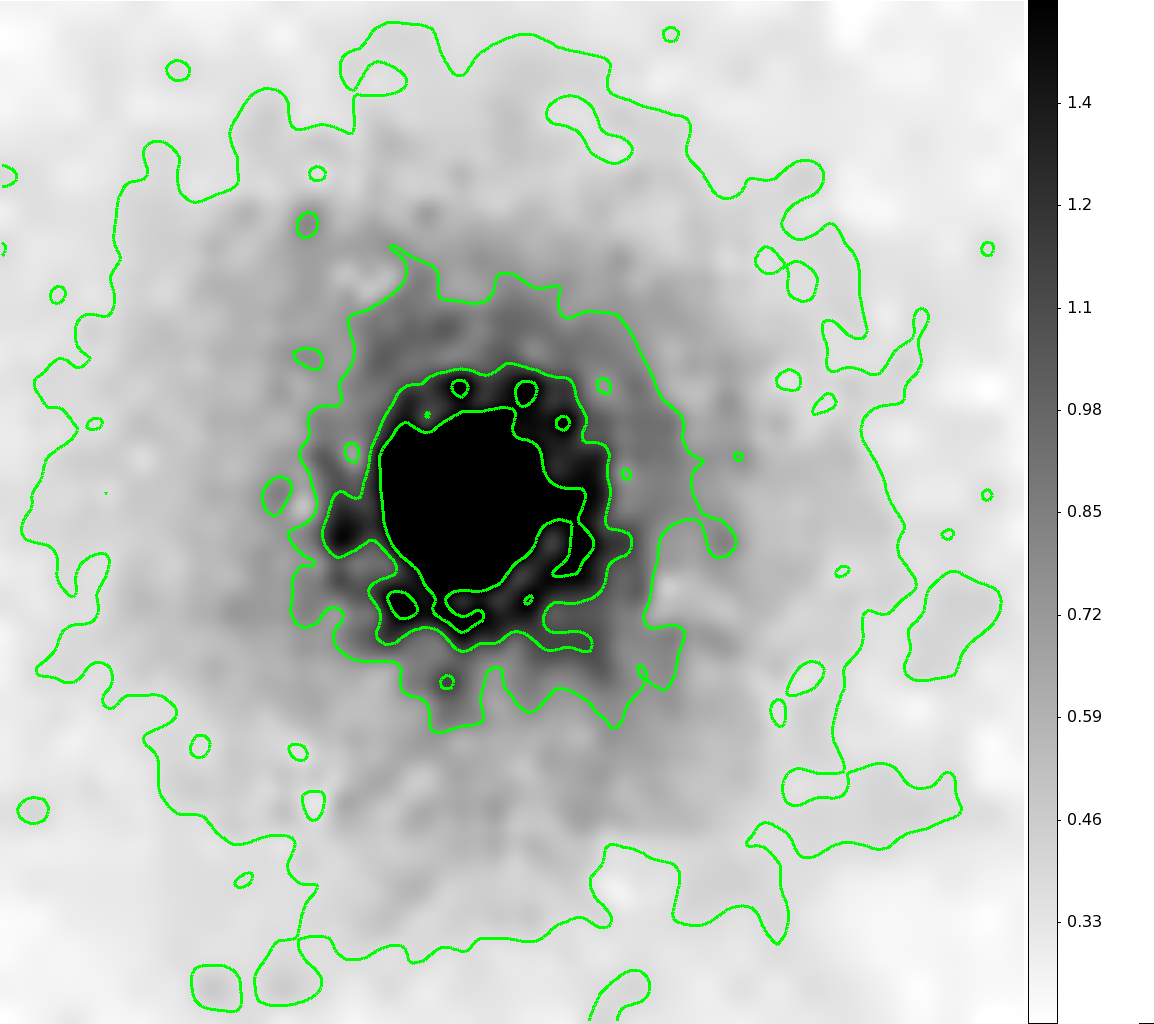} \\
		\includegraphics[width=0.45\textwidth]{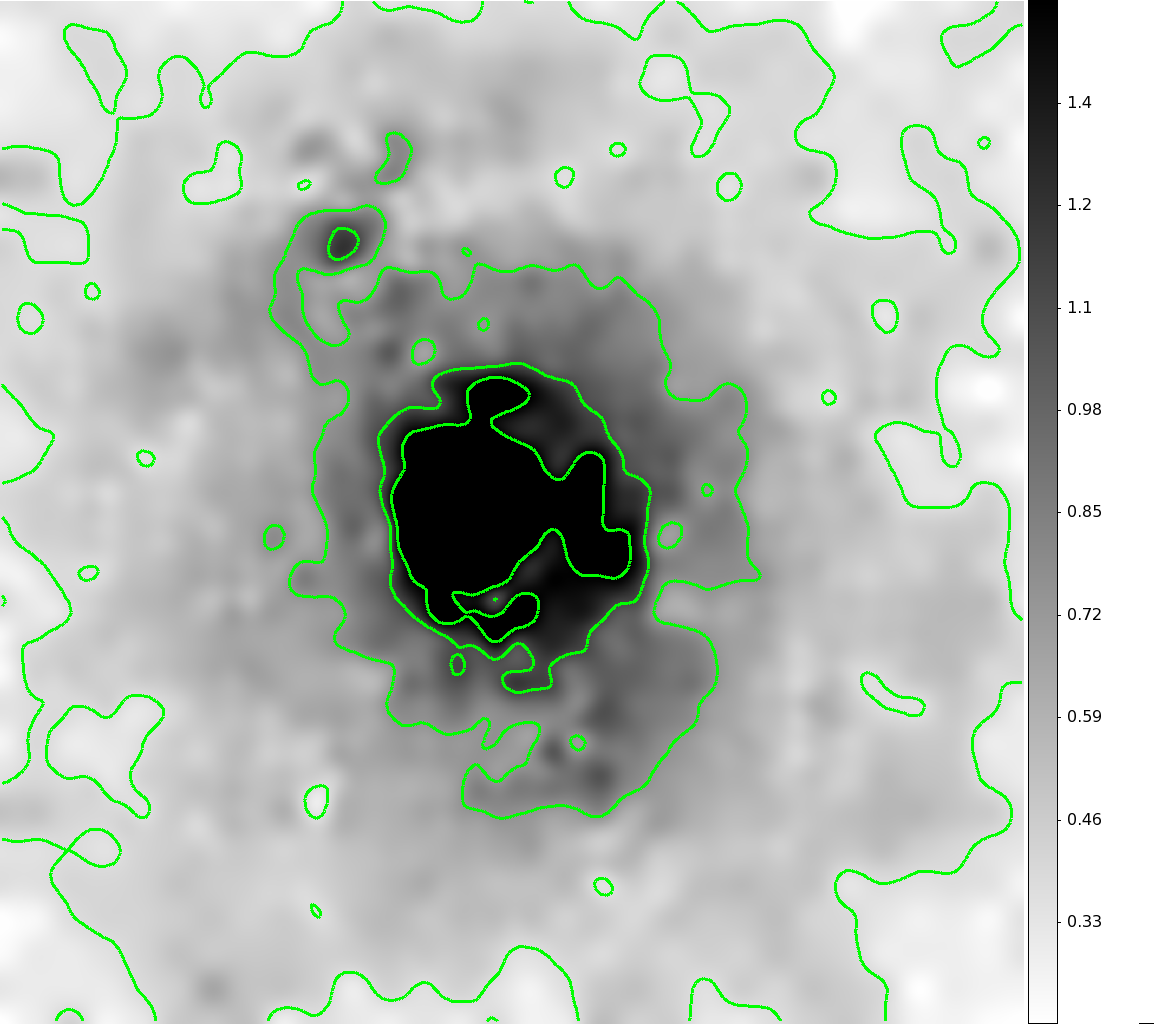} \\
	\end{tabular}		
	\caption{With flexion measurements we detect substructure in the mass distribution of Simulation 1.  From top to bottom, the input convergence (surface mass density) map, the convergence map reconstructed without flexion, and the map reconstructed including flexion data.  The innermost contour levels are the same in all three cases beginning at a convergence of 1.5 and decrease by 0.375 with each step, for a source at infinite distance.}\label{f1}
\end{figure}

Figure \ref{f1} shows a comparison of the input mass distribution, the reconstruction without flexion, and the reconstruction with flexion.  In both the without-flexion reconstruction and in the with-flexion case, the strong lensing data is extremely well fit by the reconstruction, essentially perfectly reproducing the multiple image positions for each of the sources.  The final weak lensing penalty function values in each reconstruction are very similar as well.  Both reconstructions fit the strong and weak lensing data equally well within statistical uncertainty (we discuss the statistics of these reconstructions below).

However, the reconstruction with flexion includes additional substructure undetected in the flexion-less reconstruction.  One of the two subhaloes (to the northeast) is clearly detected by eye, and the second is apparent as an extension to the southwest.  To quantify substructure detections, we employ Source Extractor \citep[SExtractor,][]{1996A&AS..117..393B} to detect ``objects'' in our mass maps.  This is similar in some ways to detecting objects in a crowded field, though with an unusually fine pixel scale.  We require a large minimum area (an area of $\sim200\text{ arcsec}^2$) above the detection threshold to identify an ``object''.  This prevents SExtractor from identifying the small, local variations in the convergence map due to the mesh point spacings as individual objects.

Because cluster subhaloes are not as discrete as galaxies in a typical field and instead are a contiguous  distribution of matter with significant blending between haloes, and because estimating the ``background'' convergence level in the environment of a parent halo is problematic, we do not discount morphological evidence for substructure in our reconstructions. While the southern substructure is not individually detected by SExtractor, it instead appears as an elongation of the source in that direction. Combining both the SExtractor detections and the visible morphology we conclude a strong detection of the northern subhalo and a likely detection of the second, though it blends partially into the main halo.

\subsection{Bootstrap Error Estimation}\label{boots}
We estimate the uncertainty in our mass reconstructions via a bootstrap resampling of the weak lensing shear and flexion catalogs.  We randomly select measurements from the two catalogs uniformly, allowing entries to be selected multiple times, to form a new pair of resampled catalogs.  From these catalogs we perform a mass reconstruction using the same initial conditions model and reconstruction parameters as used in the second step reconstruction of the original dataset.  We repeat this process 100 times, each with a different realization of the resampled catalog, to generate a distribution of mass reconstructions which we then compare to the input convergence map.  We hold the strong lensing sources constant in the bootstrapping process.  This is acceptable given that the identification of multiple images in real data allow us to more securely identify the real systems that the simulated strong lensing catalogs represent and the likelihood of mistakenly including interloper galaxies which are not strongly lensed is relatively low.

\begin{figure*}
	\begin{tabular}{cc}
		\includegraphics[width=0.45\textwidth]{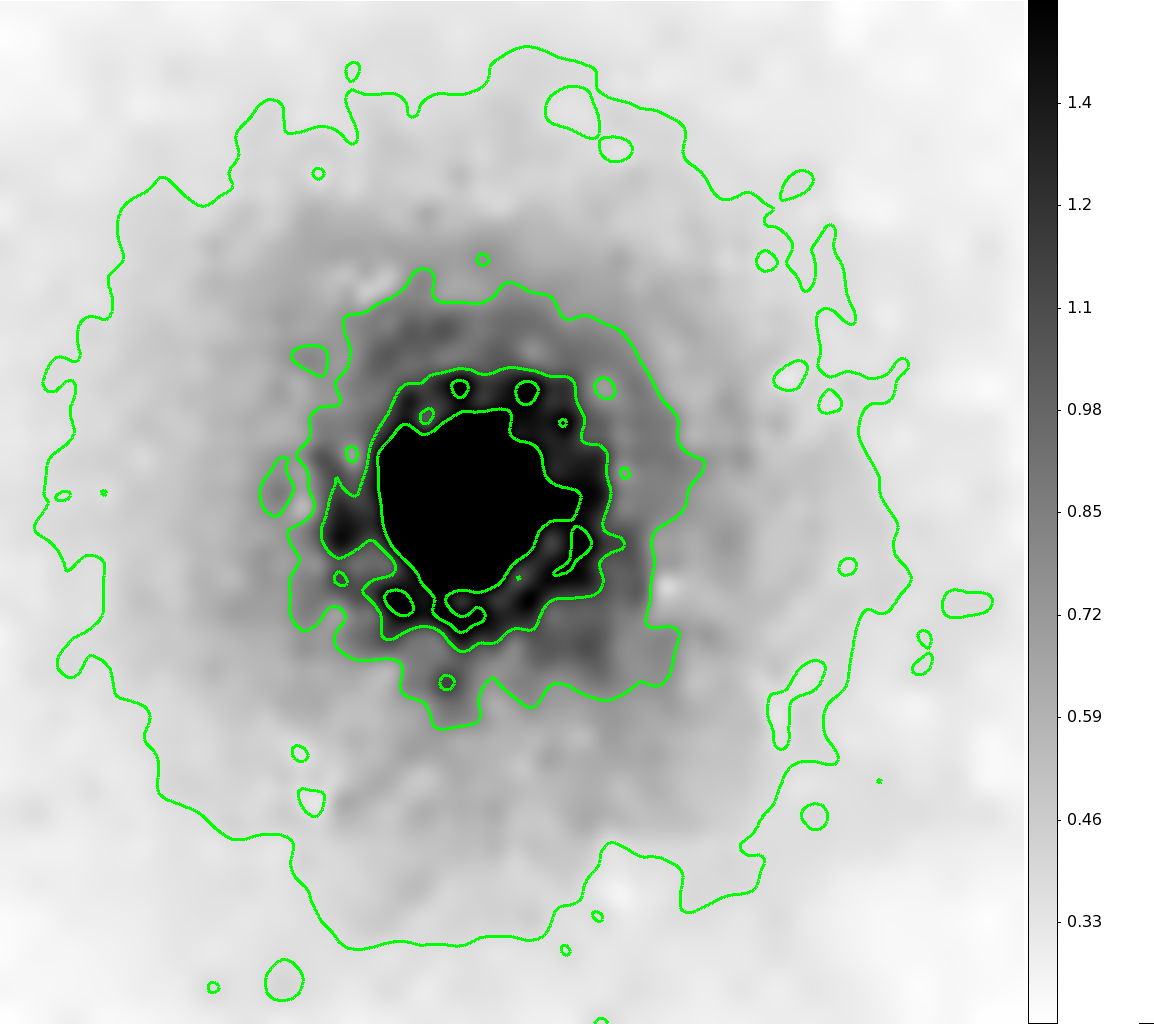} &
		\includegraphics[width=0.45\textwidth]{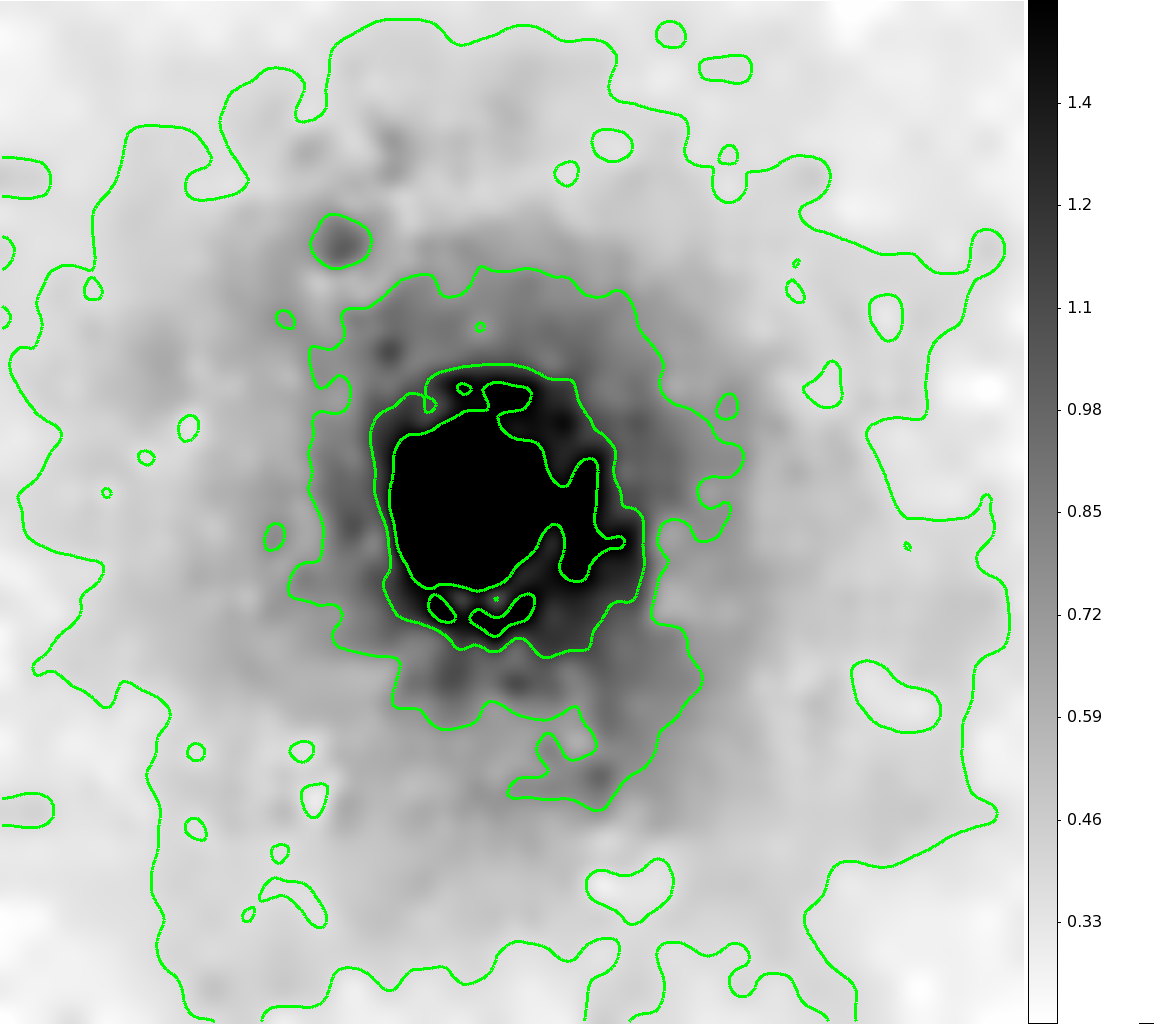} \\
		\includegraphics[width=0.45\textwidth]{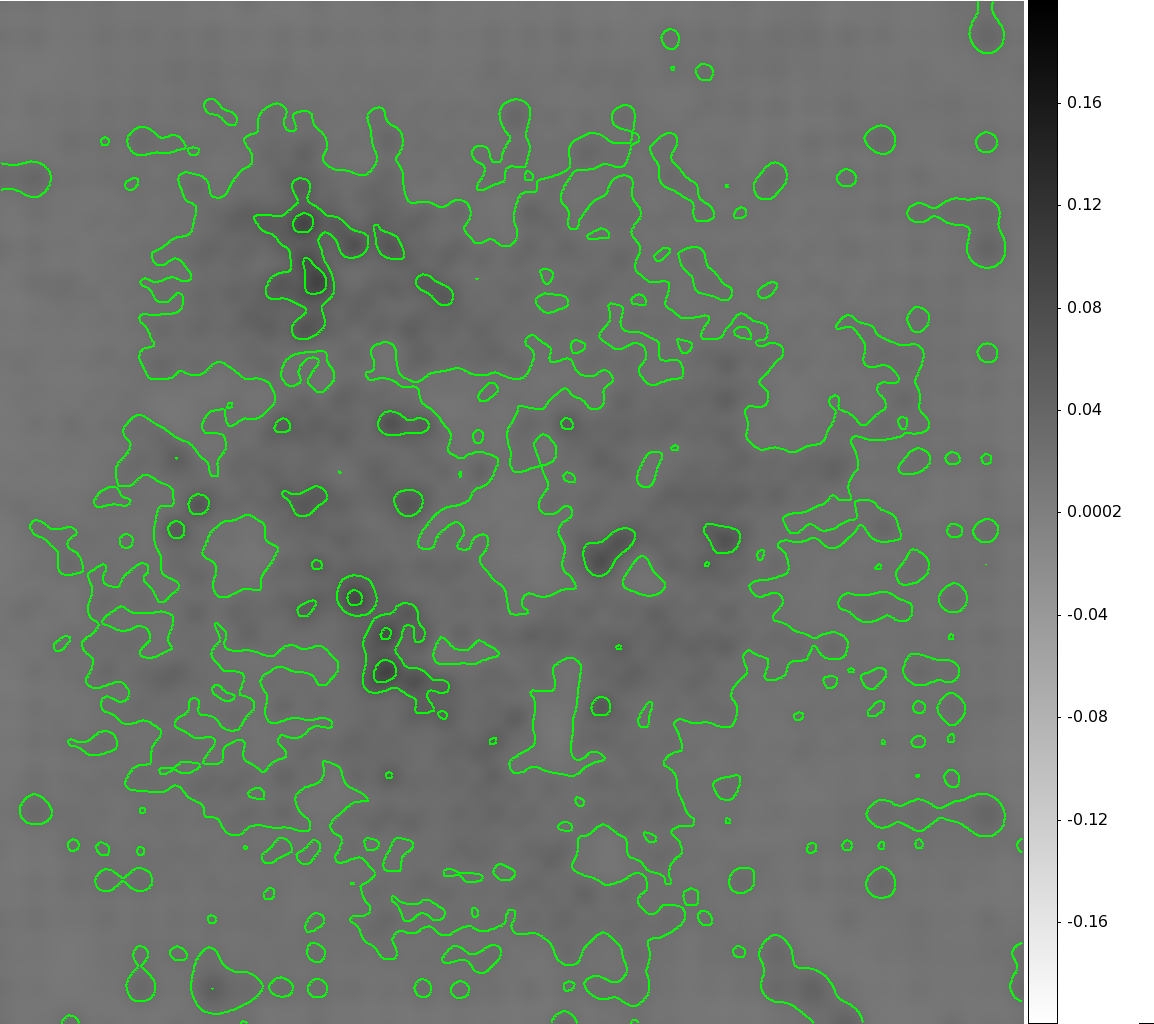} &
		\includegraphics[width=0.45\textwidth]{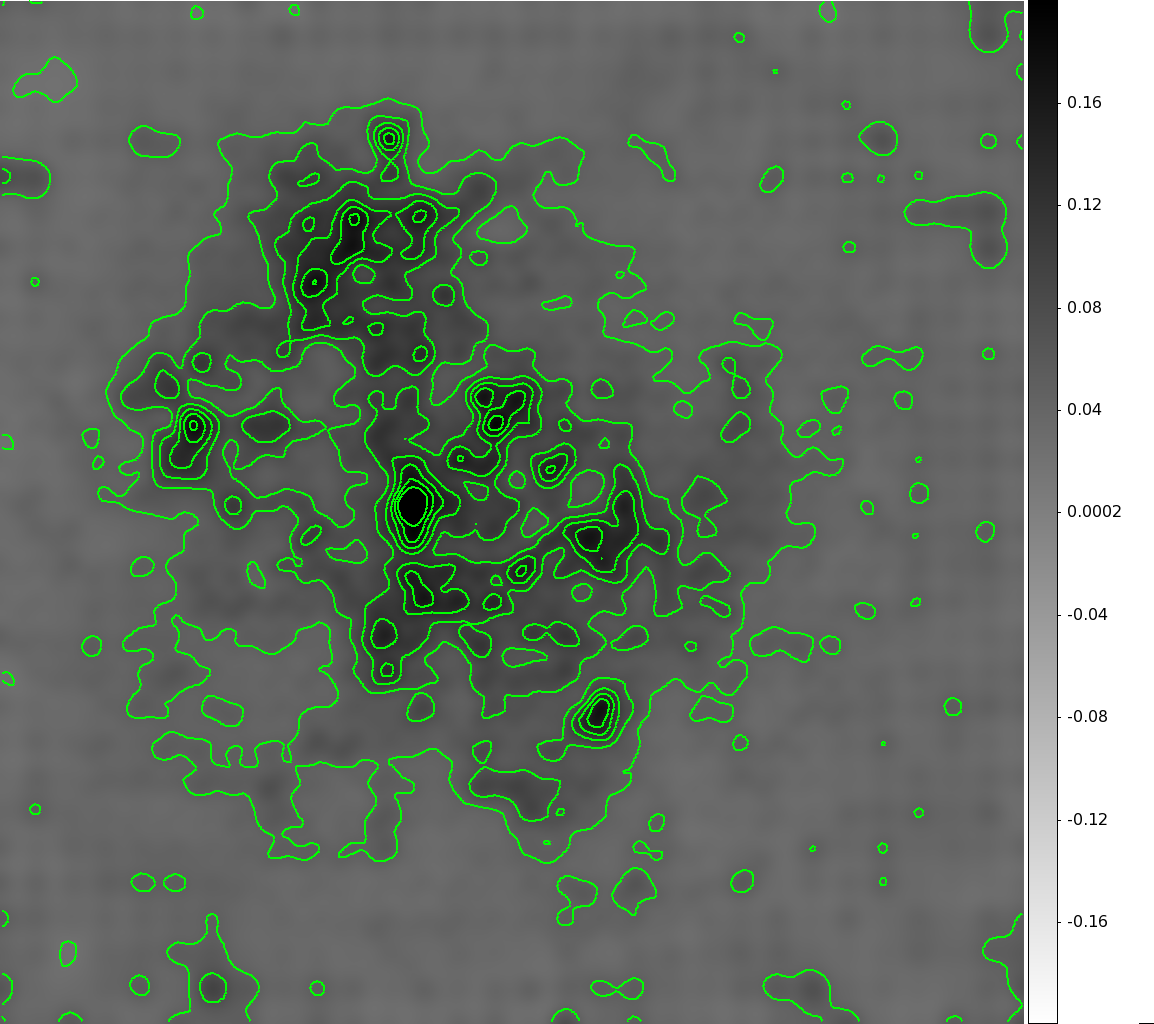} \\
	\end{tabular}
	\caption{Flexion-detected substructure is robust to catalog resampling.  On the left, the average convergence map for the set of reconstructions using the bootstrapped catalogs (above) and the RMS variation about that average (below) for reconstructions of data from simulation 1 without flexion.  On the right, the same two plots for simulation 1 reconstructions including flexion.  Contours for the top two plots are as in Figure \ref{f1}.  For the bottom two plots, the innermost contours begin at 0.2 and decrease by 0.025 with each step, also assuming a source at infinite distance.}\label{f2}
\end{figure*}

The bootstrap reconstructions produce a large data-cube of information, with a distribution of model values for each of the lensing fields: $\psi$, $\alpha_i$, $\kappa$, $\gamma_i$, $\fflex_i$, and $\gflex_i$.  Ten in all, eleven if the magnification $\mu$ is included. To characterize these distributions in a comprehensible way, we produce an average convergence map and the RMS variation of the convergence in the 100 reconstructions, which is displayed in Figure \ref{f2}.  These maps show that the substructure detected with flexion is significant relative to the scale of statistical variations in the catalog data.

\begin{figure*}
	\begin{tabular}{cc}
		\includegraphics[width=0.45\textwidth]{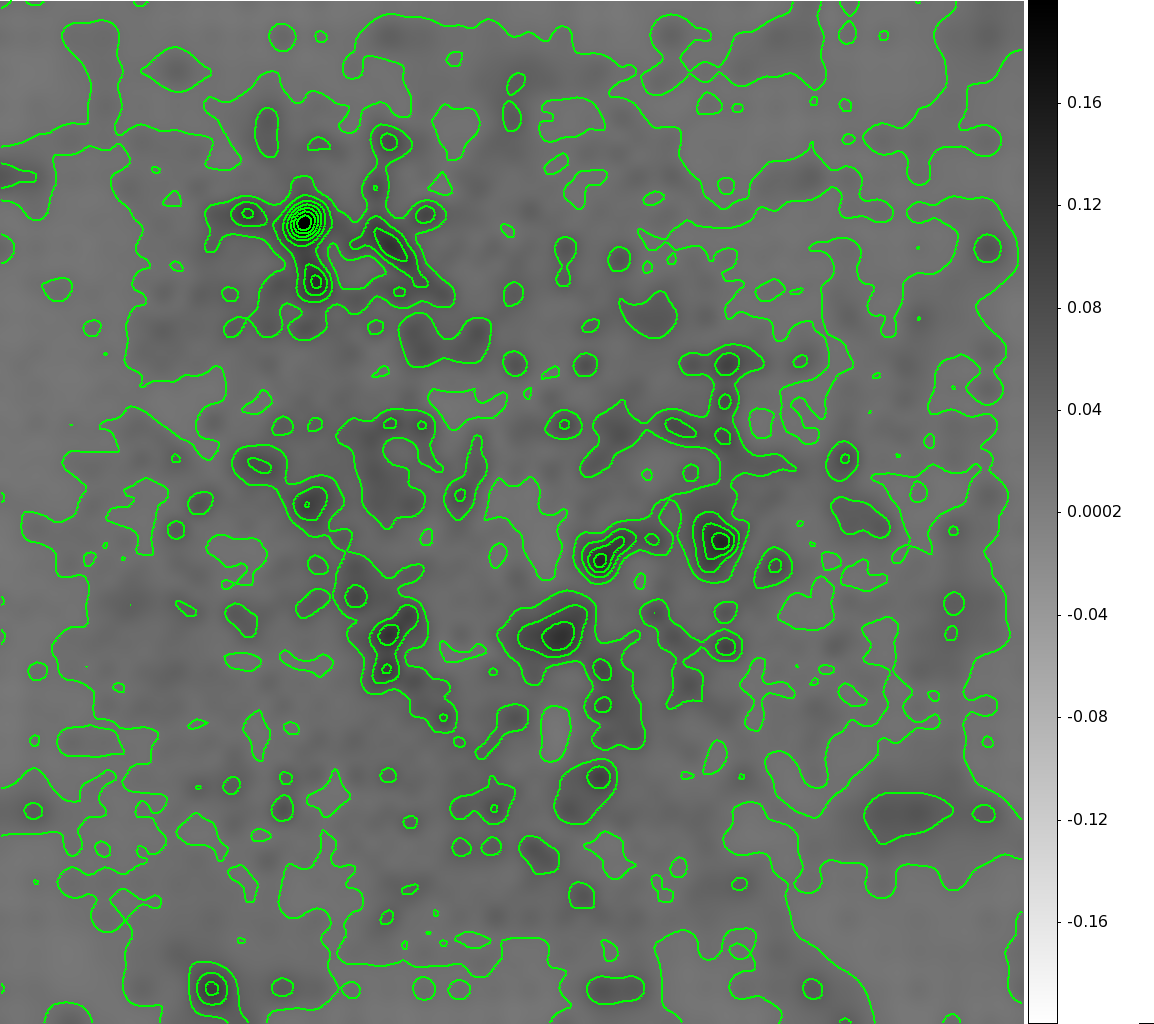} &
		\includegraphics[width=0.45\textwidth]{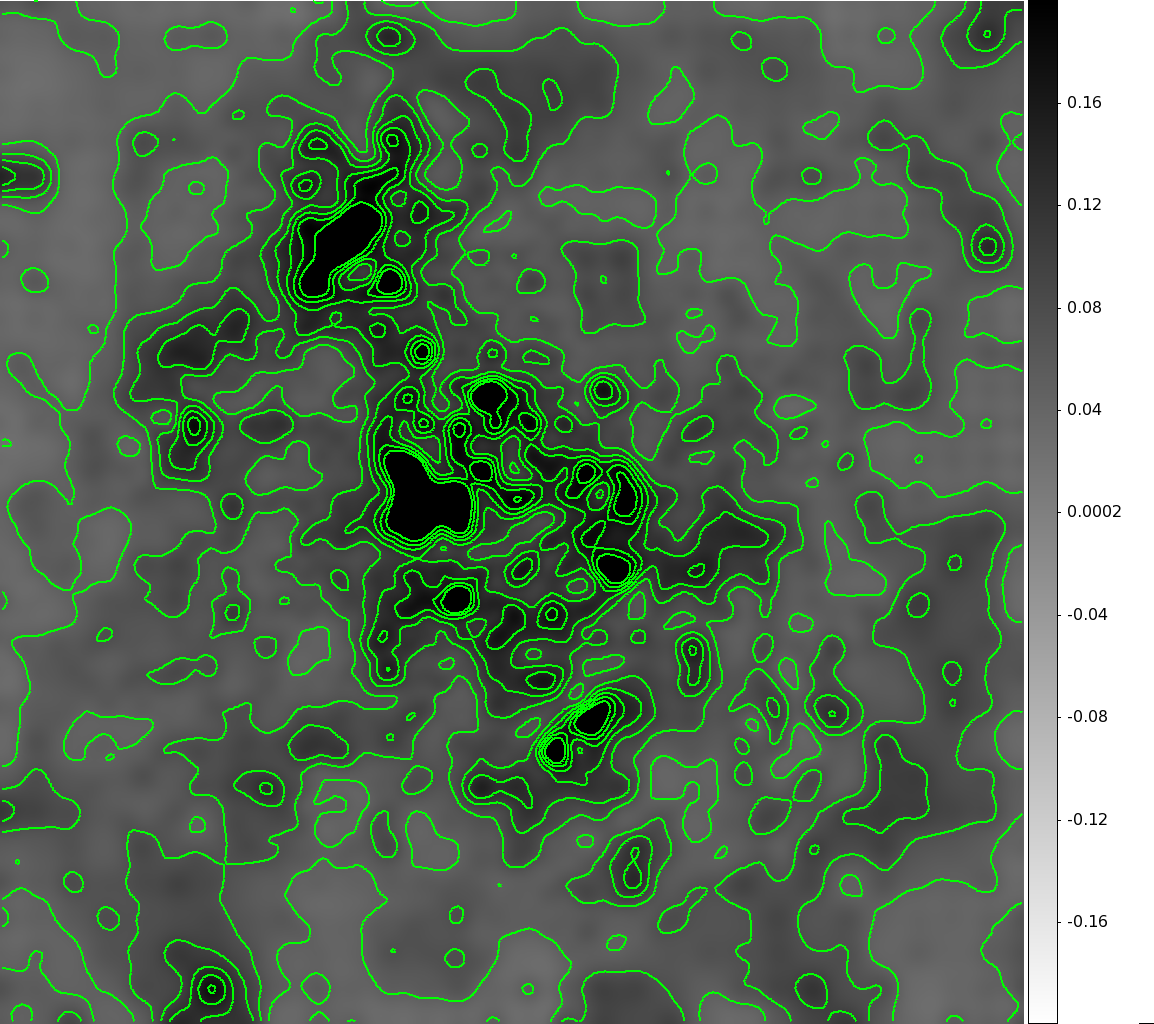} \\
		\includegraphics[width=0.45\textwidth]{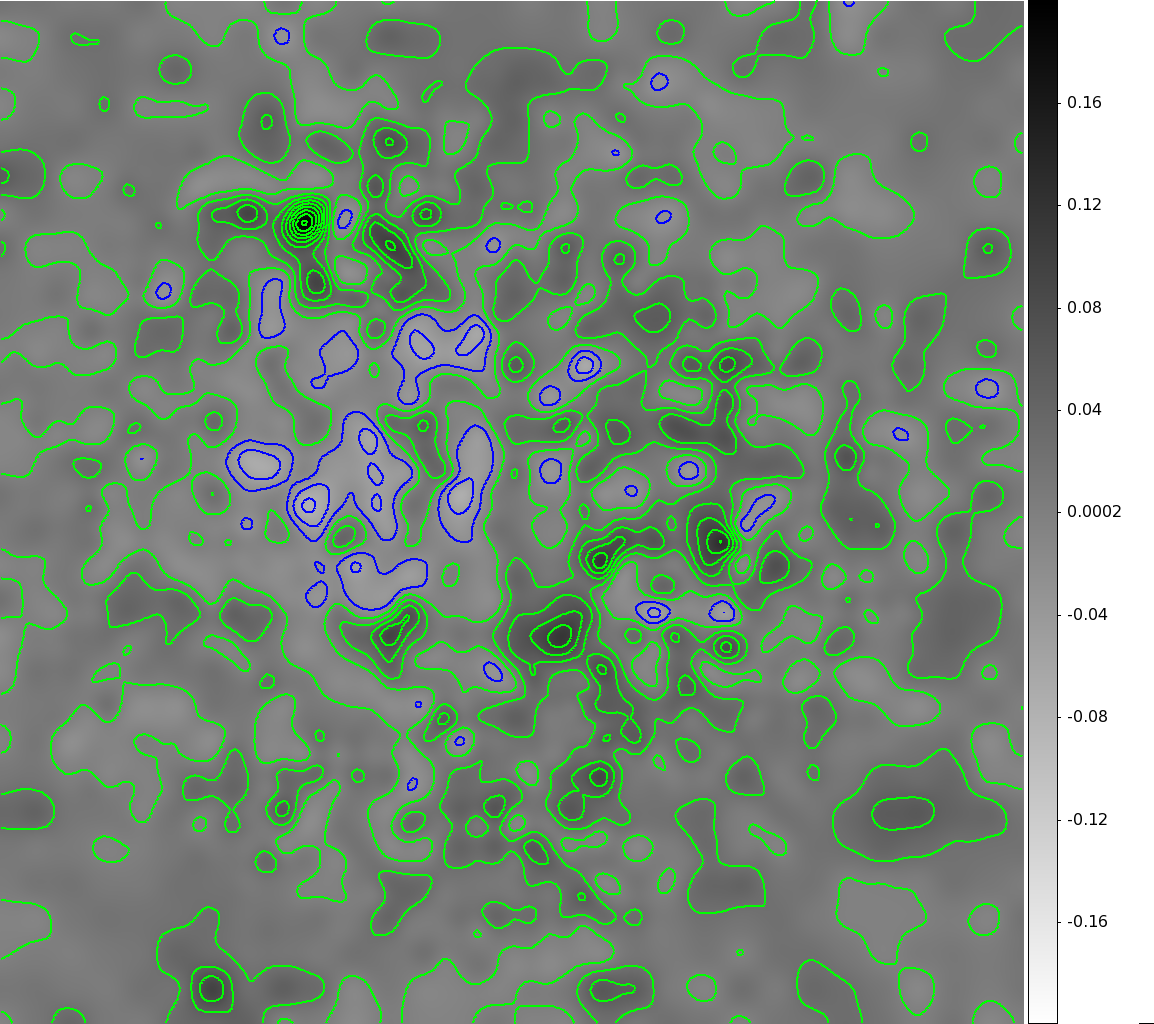} &
		\includegraphics[width=0.45\textwidth]{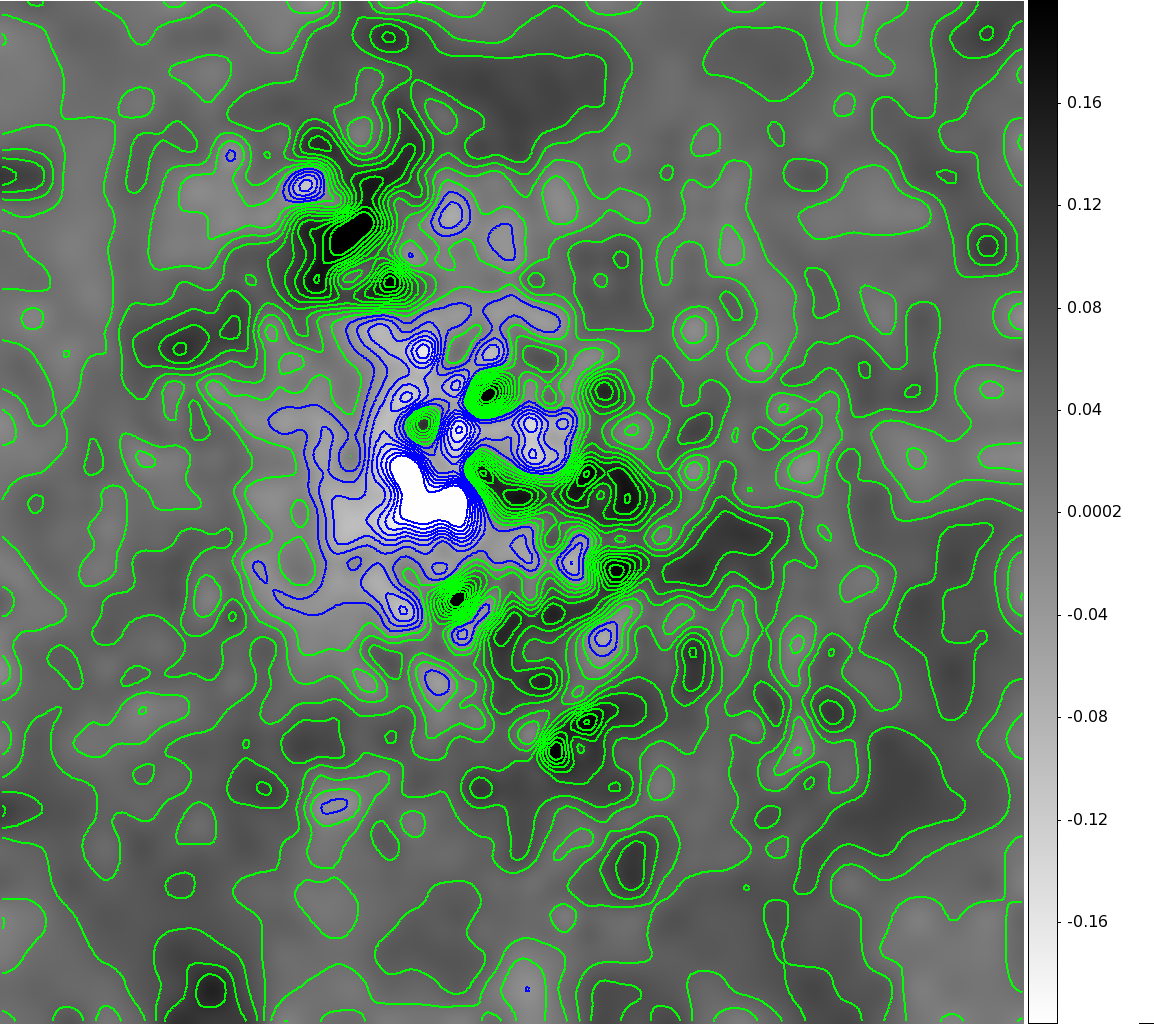} \\
	\end{tabular}
	\caption{The RMS variation of the bootstraps about the primary reconstruction (top) and the difference between the primary and the average bootstrapped reconstructions (bottom) for simulation 1.  The left column is without flexion, the right with flexion.  Green contours indicate positive values, with the innermost contours at 0.2 decreasing in steps of 0.025 to zero.  Blue contours indicate negative values, with the innermost value at -0.2 and the same step size.}\label{f3}
\end{figure*}

We also map both the RMS variation of the bootstrapped convergence maps relative to the primary reconstruction, and also the difference between the primary and average convergence maps, presented in Figure \ref{f3}.  These give us a metric of bias in the primary reconstruction, i.e., how much of an outlier is the primary reconstruction.  Were we to see a large amount of bias in these maps, meaning a large difference between the RMS variation about the average and the RMS variation about the primary, or were we to see a large difference between the primary and the average convergence maps, this would suggest that the primary reconstruction were a statistical outlier rather than a robust minimum solution to reconstructing the mass distribution.  We do not see any evidence that this is the case.

We see the expected small number of outliers among the reconstructions, but overall the bootstrapped catalog reconstructions should not vary drastically from the primary reconstruction if they are valid estimators of the error in the reconstruction.  A few reconstructions ($\lesssim5$) have hugely discrepant penalty function values, universally due to a poor fit to the the strong lensing data and thus an incorrectly located critical curve.  These outliers are easily removed and discarded from our statistical analyses, though leaving them in does not significantly affect the results as the lensing field maps, such as the convergence, are not catastrophically deviant from the primary solution even with these outlier solutions since they depend less on the exact position of the critical curve.

The reconstruction method is computationally intensive, requiring the inversion of large sparse matrices used for the finite-differences calculation of the gradient of the penalty function with respect to the lensing potential values on the mesh, so we perform this bootstrap analysis on only one of our simulations (Simulation 1), both with and without flexion measurements included, to estimate the amount of variation to expect due to the statistical uncertainty in our lensing field estimator measurements.  As with real data, the variation depends not only on the accuracy and density of the weak lensing and flexion estimators across the field, but also on the amount of specific constraint provided by the particular set of strong lensing images available.  The number and, importantly, the location of the multiple images have an effect on the spatially varying amount of freedom that the mass model can change within the statistical errors of the measurements.  A more full statistical analysis of the cluster lensing mass reconstruction made including flexion allowing a broader variation of possible halo and subhalo configurations is a large undertaking, and beyond the scope of this work.  However we can use our bootstraps to begin to estimate the errors in our mass maps, and compare the difference between the input mass distribution and the reconstructed mass map to these error estimates.

Using the bootstrap-resampled catalog reconstructions, we can understand the statistical variation in our mass maps due to the catalog sampling. One immediate feature to note in the non-flexion RMS map (Figure \ref{f2}) is that the variation is small and is relatively uniform across the field.  This means that despite the catalog resampling, a consistent mass distribution is preferred by the data. The input subhaloes are larger than any of the statistical variations by quite a large margin, the non-detection of the subhaloes is not due to noise or a statistical fluctuation.  Instead, it is due to incomplete information in the lensing inputs, namely the lack of flexion information measuring the mass and shear gradients.

The RMS map for the flexion reconstruction is less uniform than the non-flexion RMS map, which can be understood considering how flexion measurements inform the mass reconstruction.  Because the flexion field is sourced very locally by the gradient of the mass distribution, and since the flexion measurement is not dominated by statistical noise in the same way that shear measurements are, excluding certain individual flexion measurements from the catalog can have a more significant effect on the resulting mass map than happens with weak lensing shear measurements that, by their nature, must be correlated together to extract the lensing signal.  Thus we expect that bootstrapping the flexion measurement catalogs in addition to the shear catalogs will increase the RMS values near where the flexion measurements are most informative and constraining.  This includes the locations of substructures, as can be seen in the RMS maps in Figures \ref{f2}-\ref{f3}, but also where the flexion field is strong but reliable measurements are relatively sparse (near the main halo center) as well as where flexion measurements are particularly near a strong lensing image.  The RMS map shows that neither substructure detection can be attributed to the statistics of the flexion information, with one certain detection (north) and a second, less significant detection (south).  

As noted above in comparing the average reconstruction to the primary reconstruction, both with and without flexion, we see no significant overall bias (Figure \ref{f3}).  For reconstructions with flexion, the average reconstruction underestimates the mass in the substructures as is expected since the bootstrap reconstructions include catalogs where some of the flexion measurements nearest the substructures are not included which will reduce the lensing signal from that substructure.  The features in these maps are what would be expected given the information available (and not available) to each reconstruction.

These results, the primary mass reconstruction together with the properties of the reconstructions from bootstrap resampled catalogs, show that the inclusion of flexion measurements can detect substructures which are otherwise undetectable in strong lensing and weak lensing shear data alone.  Furthermore, these detections are statistically significant and robust to the variations of bootstrapped catalogs, i.e., they do not depend solely on a few fortuitously-located images, though there is a more marked dependence of the resulting mass map on the location of the flexion measurements relative to any substructures than is seen for weak lensing shear measurements.

%%%%%%%%%%%%%%%%%%%%%%%%%%%
\subsection{Simulations 2 through 6}
\begin{figure*}
	\begin{tabular}{cc}
		\includegraphics[width=0.45\textwidth]{sim00_true_kappa_inf.png} &
		\includegraphics[width=0.45\textwidth]{sim00_fly_kappa_inf_smth.png} \\
		\includegraphics[width=0.45\textwidth]{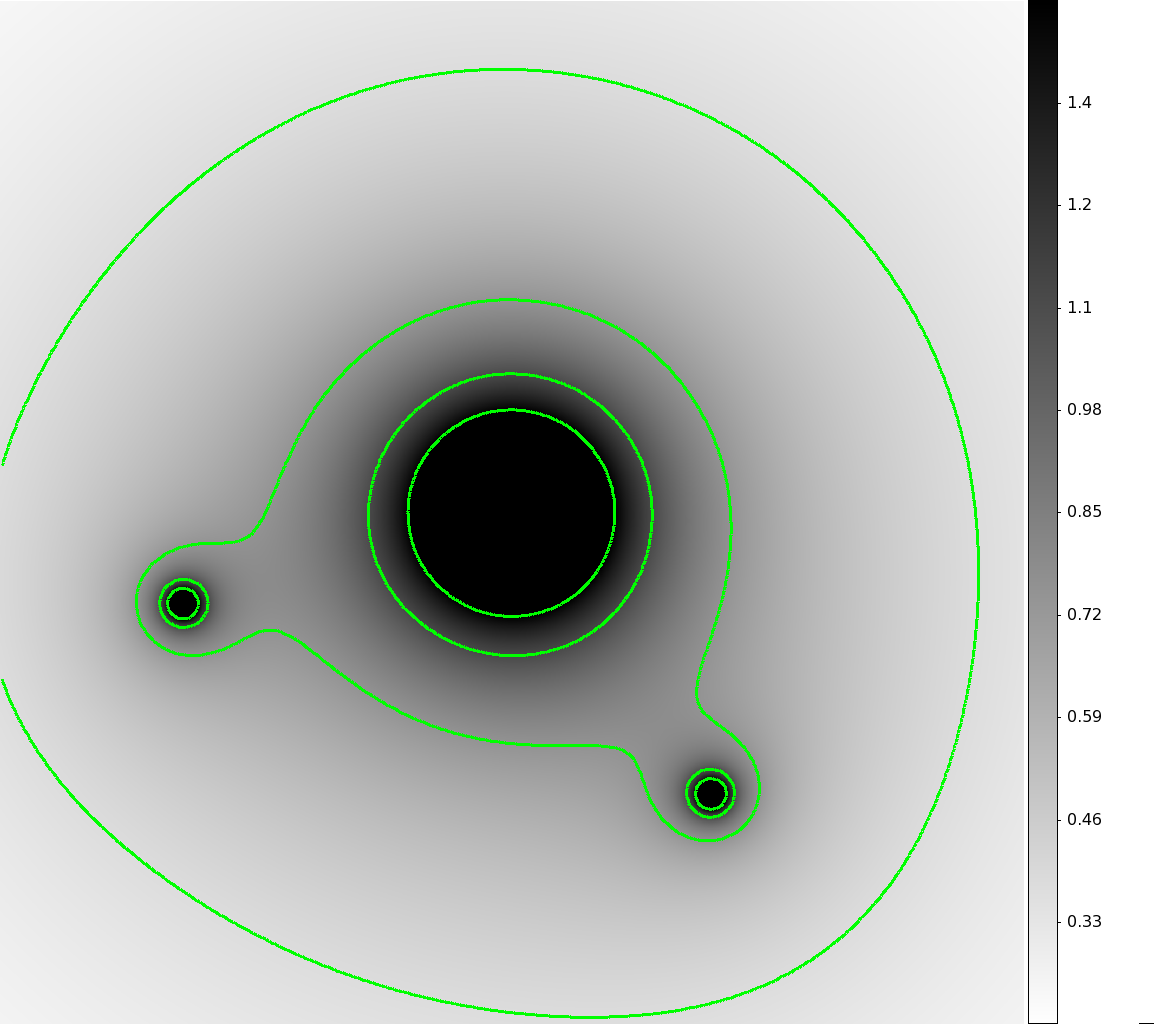}  &
		\includegraphics[width=0.45\textwidth]{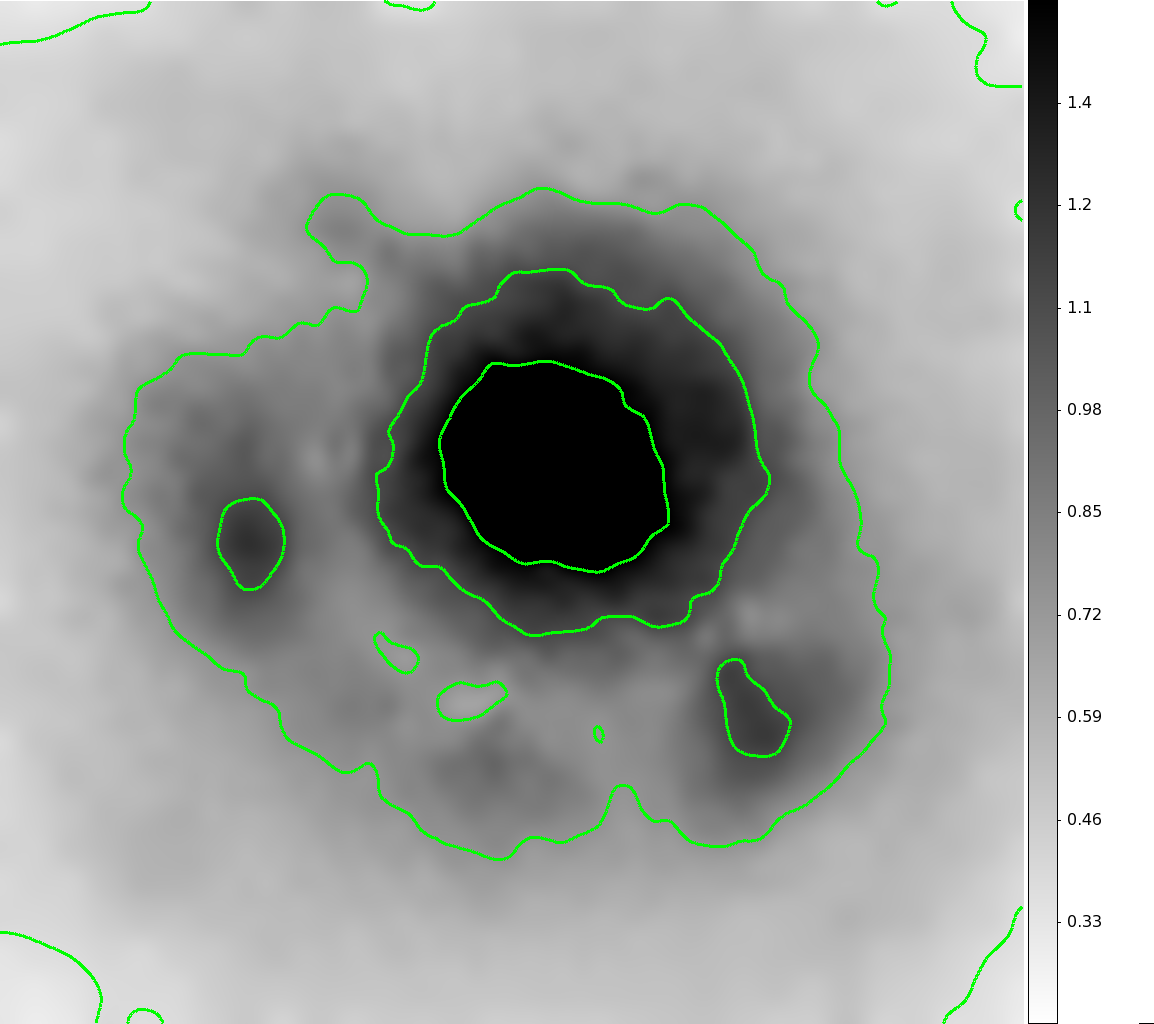}\\
	\end{tabular}
	\caption{A comparison of input mass distributions (left) and the reconstructions including flexion (right) for simulations 1 (top) and 2 (bottom).  Contours are as in Figure \ref{f1}. In these plots, and in Figures \ref{f5}-\ref{f6}, the reconstruction field displayed is centered on the strong lensing image centroid, and therefore are slightly offset from the true halo center.}\label{f4}
\end{figure*}

\begin{figure*}
	\begin{tabular}{cc}
		\includegraphics[width=0.45\textwidth]{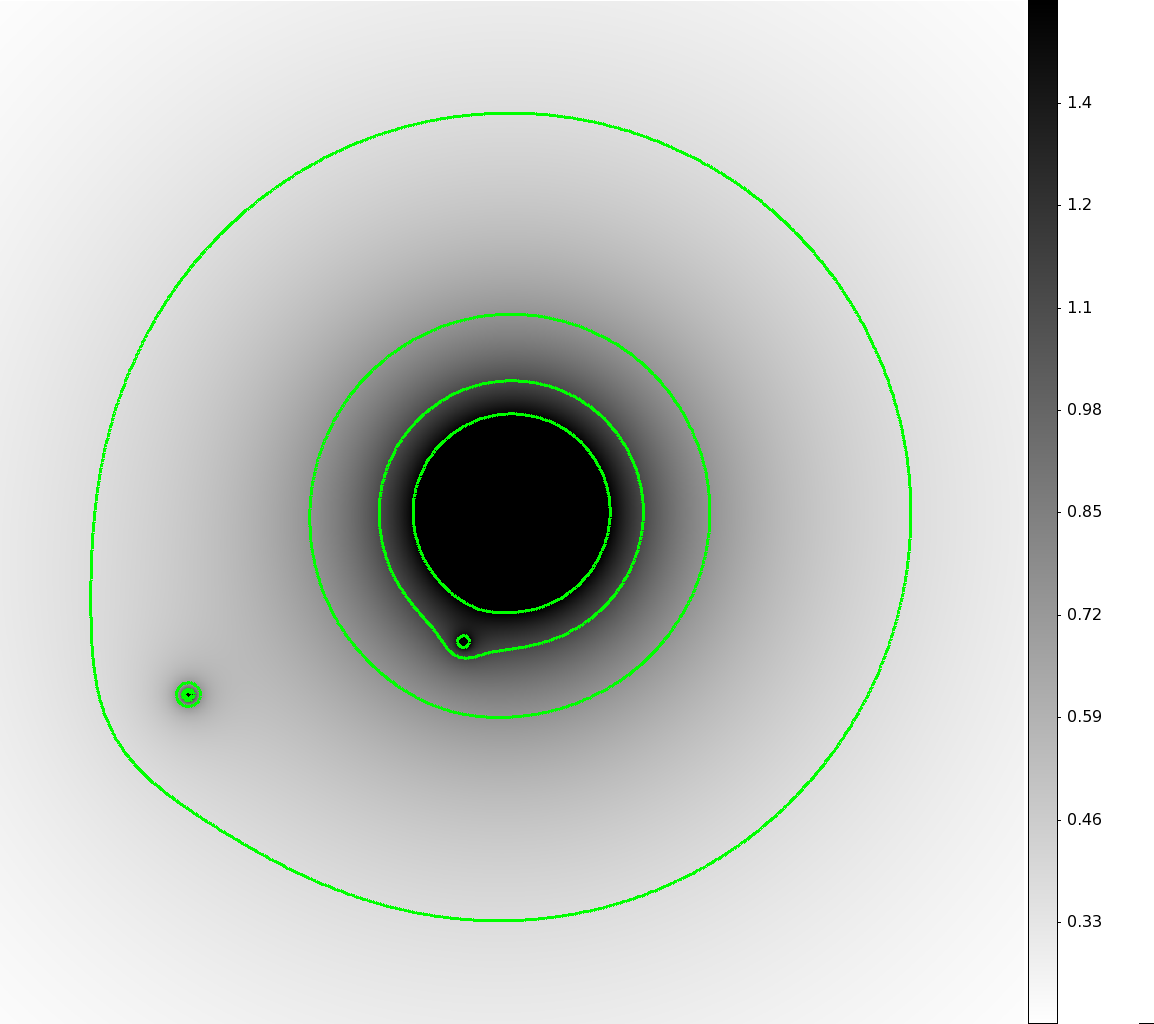} &
		\includegraphics[width=0.45\textwidth]{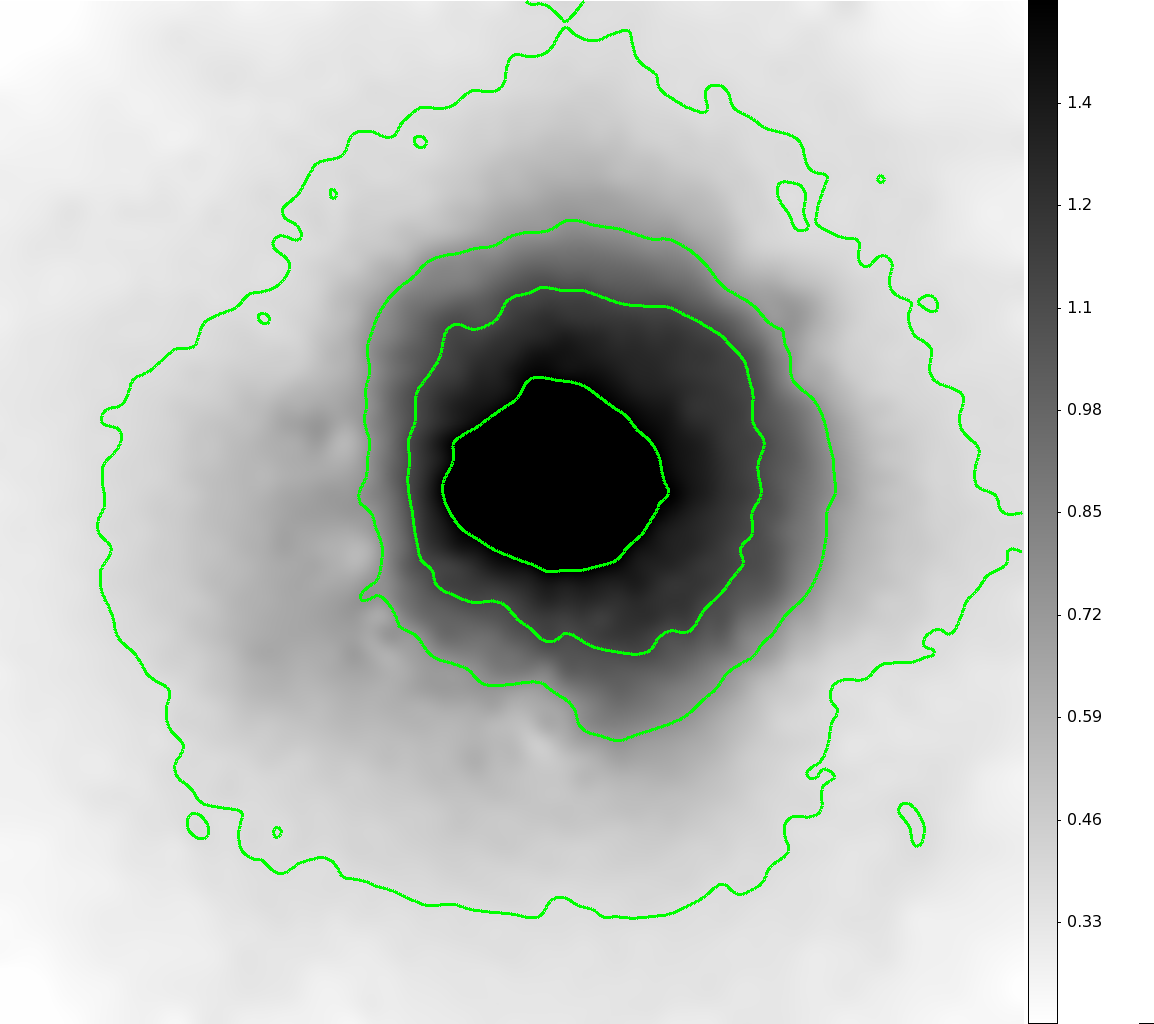} \\
		\includegraphics[width=0.45\textwidth]{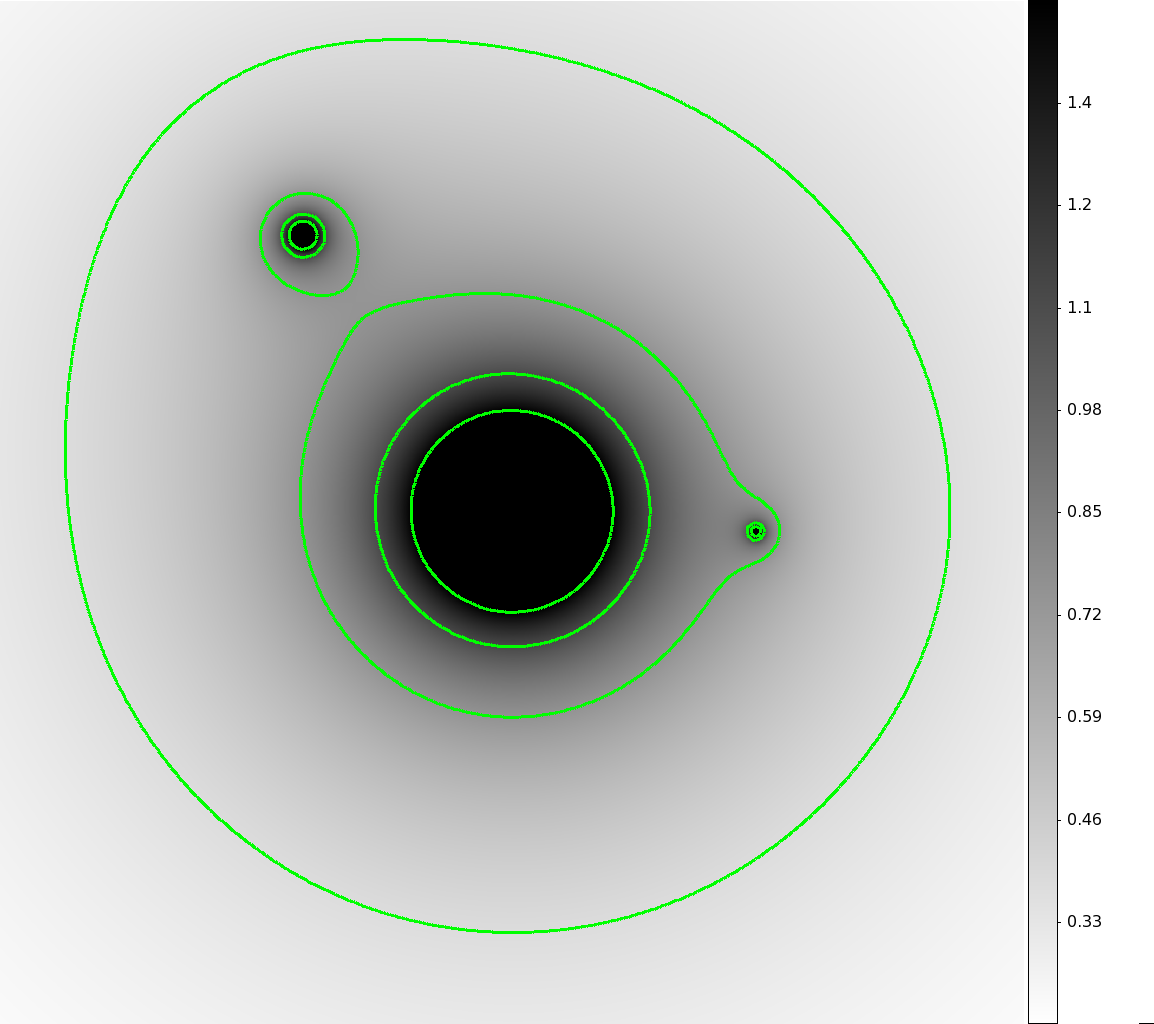} &
		\includegraphics[width=0.45\textwidth]{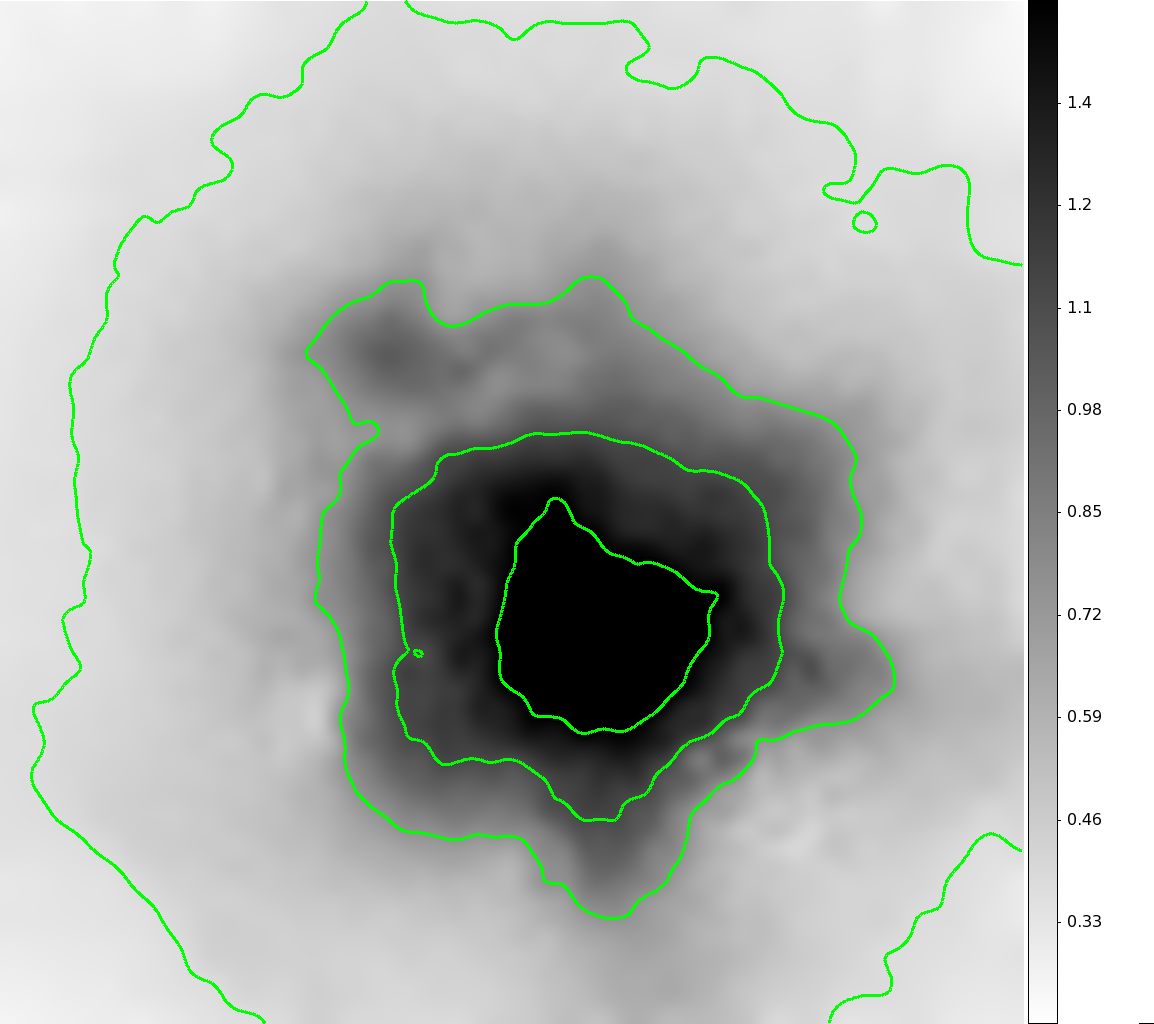} \\
	\end{tabular}
	\caption{As in Figure \ref{f4}, for simulations 3 (top) and 4 (bottom).}\label{f5}
\end{figure*}

\begin{figure*}
	\begin{tabular}{cc}
		\includegraphics[width=0.45\textwidth]{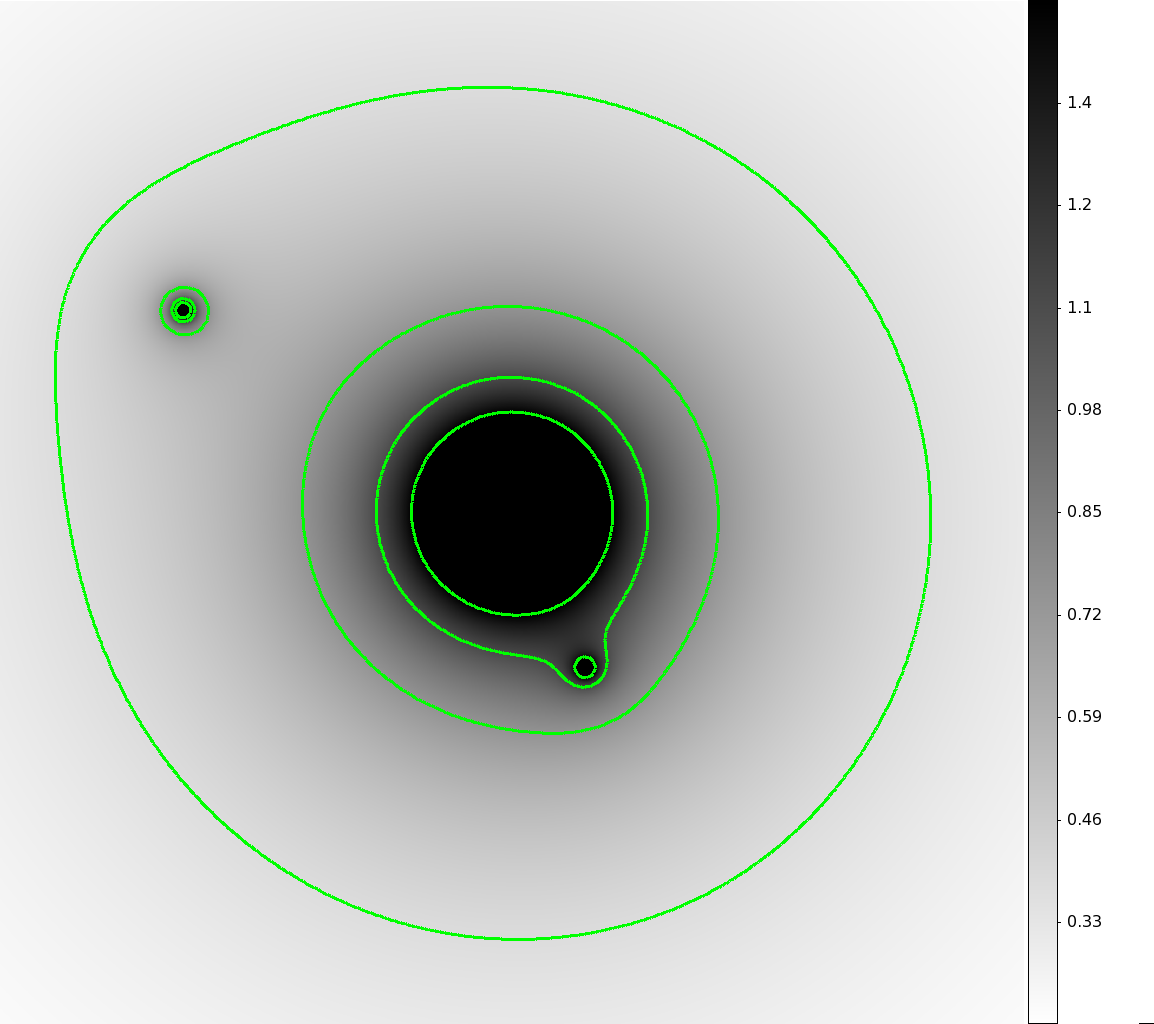} &
		\includegraphics[width=0.45\textwidth]{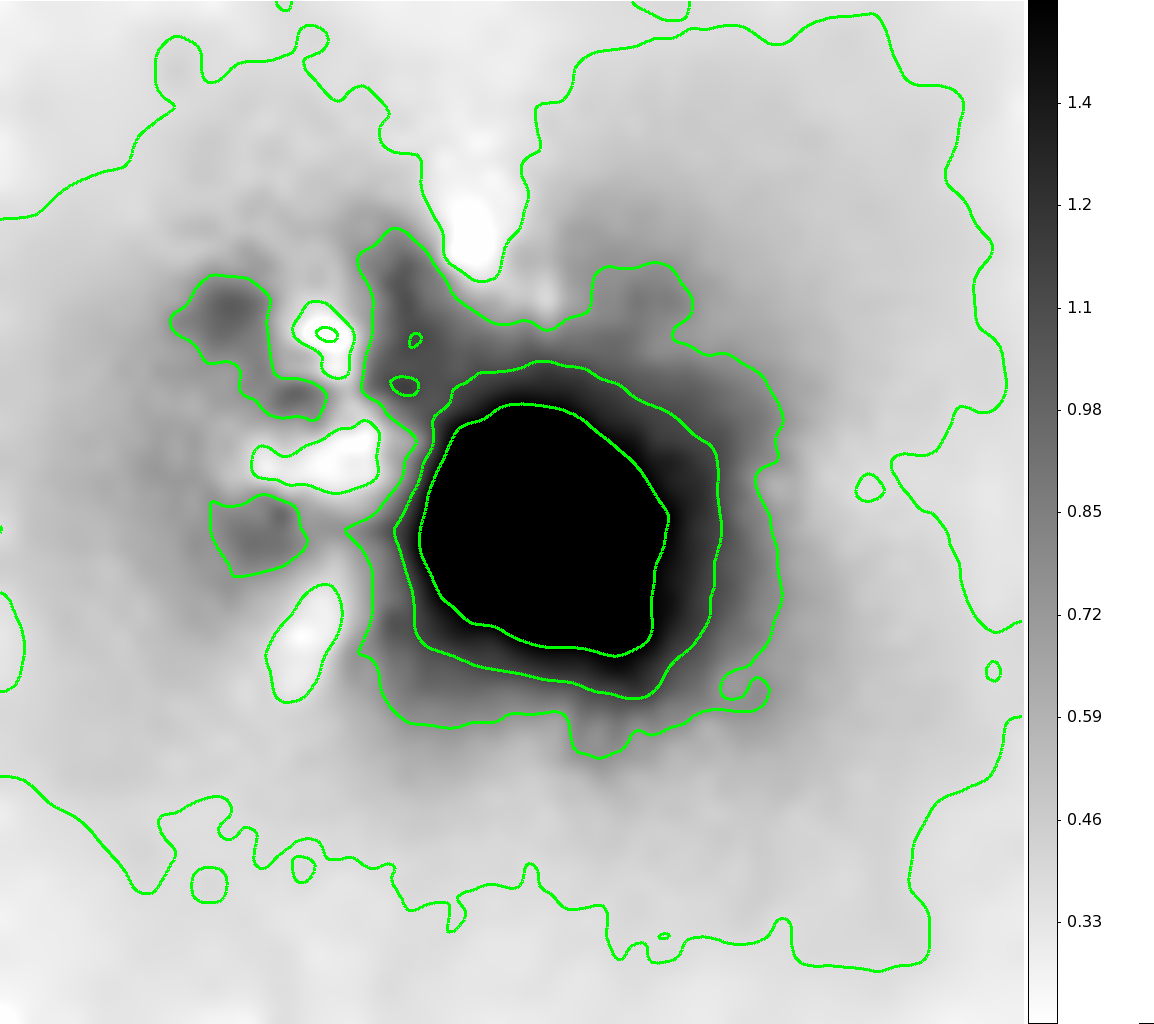} \\
		\includegraphics[width=0.45\textwidth]{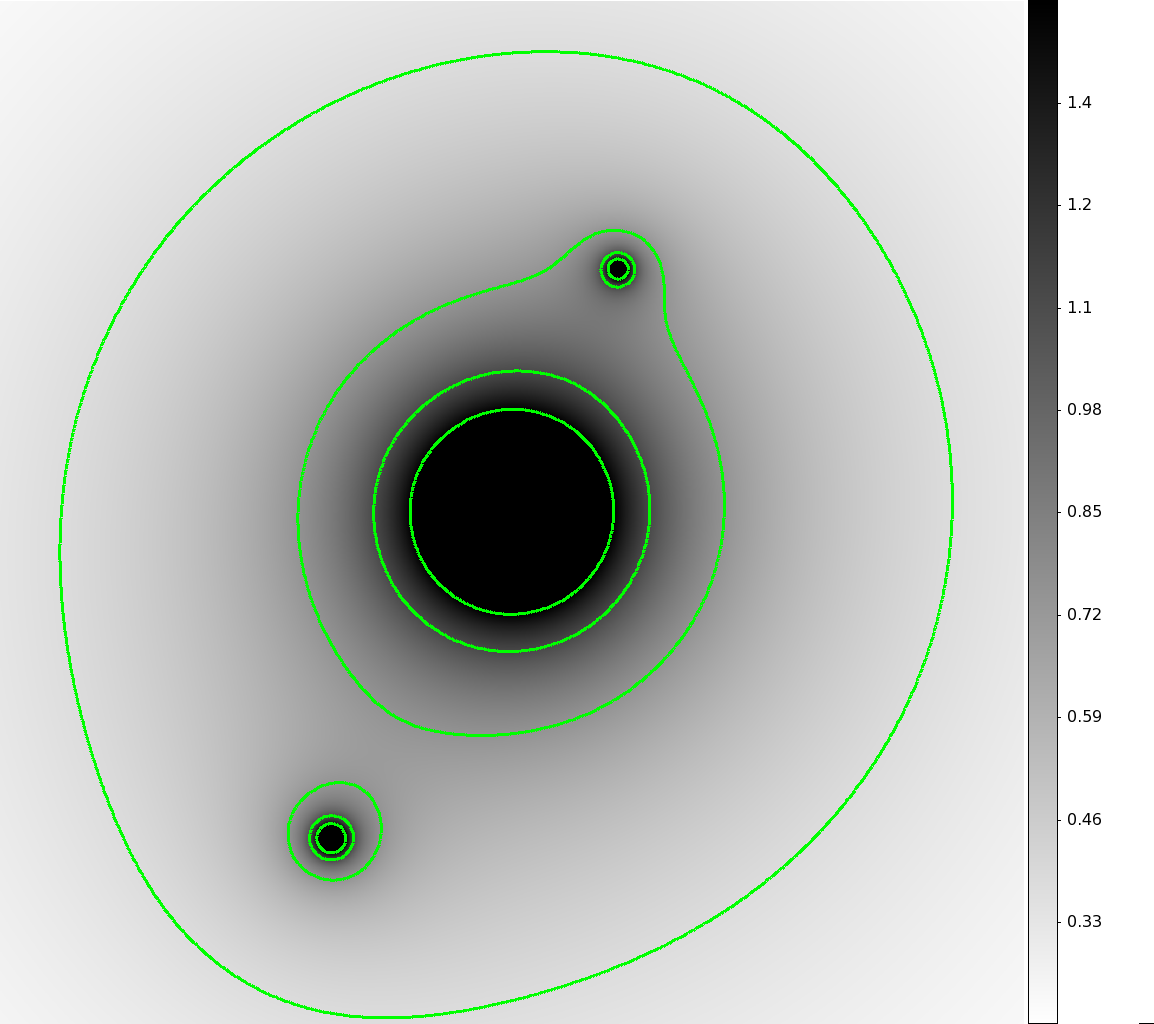} &
		\includegraphics[width=0.45\textwidth]{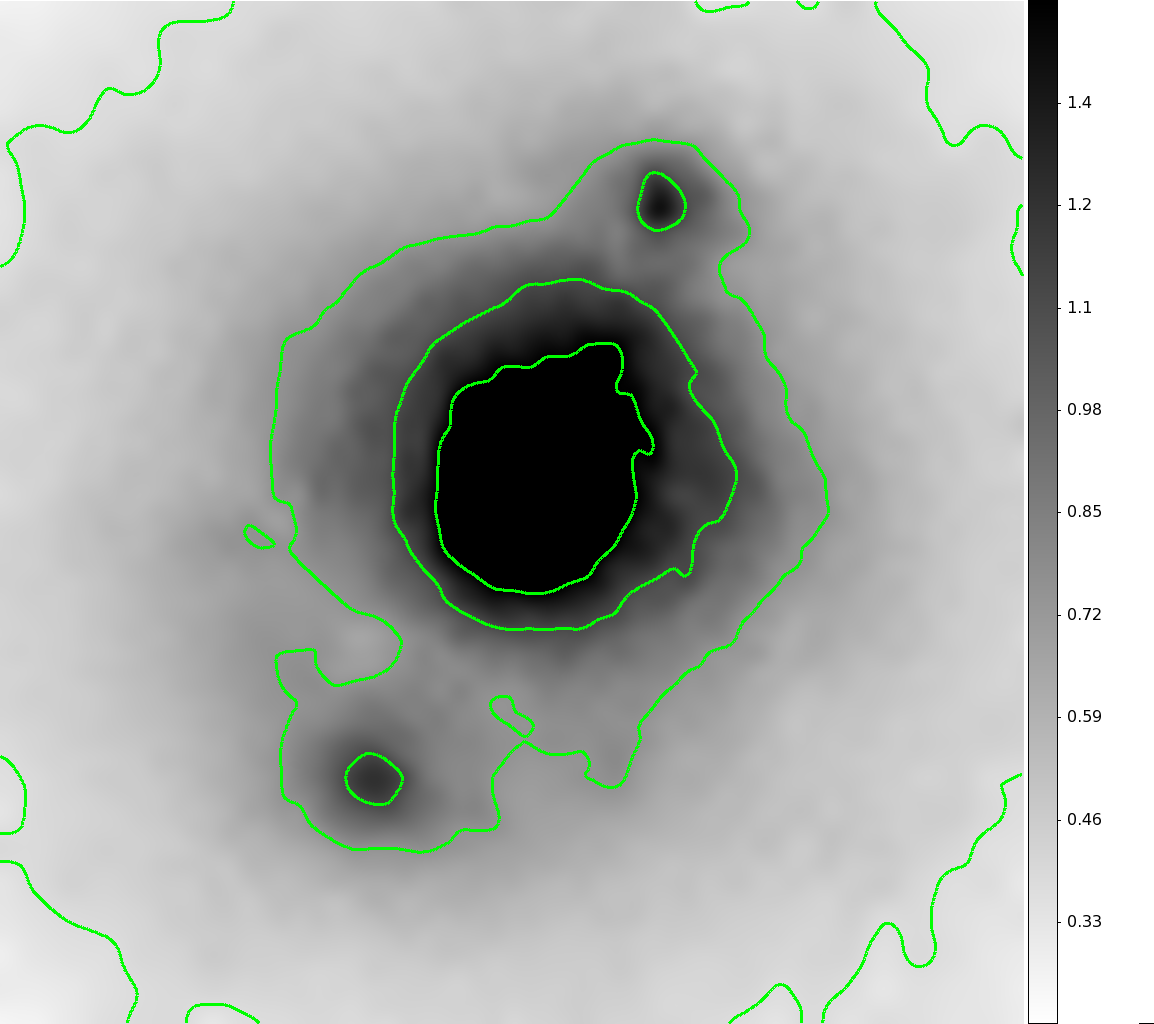} \\
	\end{tabular}
	\caption{As in Figure \ref{f4}, for simulations 5 (top) and 6 (bottom).}\label{f6}
\end{figure*}

Figures \ref{f4}, \ref{f5}, and \ref{f6} each show comparisons between the reconstructed mass distributions (on the left) and the input mass distribution (right) the eight simulations.  These, along with Table \ref{apmasstable} show the varying substructure sensitivity enhancement that including flexion yields.  Simulations 1 and 2 each have two equal mass subhaloes, though the positions differ between the simulations.  For simulations 3-6, we vary the masses of the subhaloes.

A few trends emerge from these simulations.  As would be expected, more massive substructures are reliably detected, particularly if they are well separated from the core of the main halo.  The one instance where a larger subhalo is not detected (simulation 7) is one where the subhalo is blended into the main halo.  In many configurations, this type of subhalo (in terms of mass and location) would be detectable from strong lensing data alone, as we know from the set of halo+substructure configurations we generated but rejected for this study due to their obviously substructured strong lensing observations.

For smaller substructures, the detectability decreased with distance from the main halo.  This is also expected.  The value of the non-reduced flexion fields (i.e., $\fflex$ and $\gflex$) is dominated by the gradients in the convergence and shear field whose distortions depend on the proximity to the subhalo mass distribution. The total convergence (which factors into the reduced flexion) is dominated by the main halo mass.  At larger radii from the main halo center, the total convergence falls away from unity, the flexion decreases quickly away from the subhalo, and both effects cause the measurable reduced flexion signal to decrease.  The least massive subhalo we detect (in simulation 4), though we only detect it marginally, was located quite close to the main halo center where the total convergence is much closer to unity.  More radially distant subhalos were either undetected or more massive.  As we discuss in detail in \S\ref{sec:apmass}, the inclusion of flexion does more than detect substructure and improve the reconstructed ellipticity.  Flexion also improves the overall fidelity of the mass reconstruction independent of the detectability of individual structures.

\subsection{Aperture Masses}\label{sec:apmass}
\begin{figure}
	\includegraphics[width=0.5\textwidth]{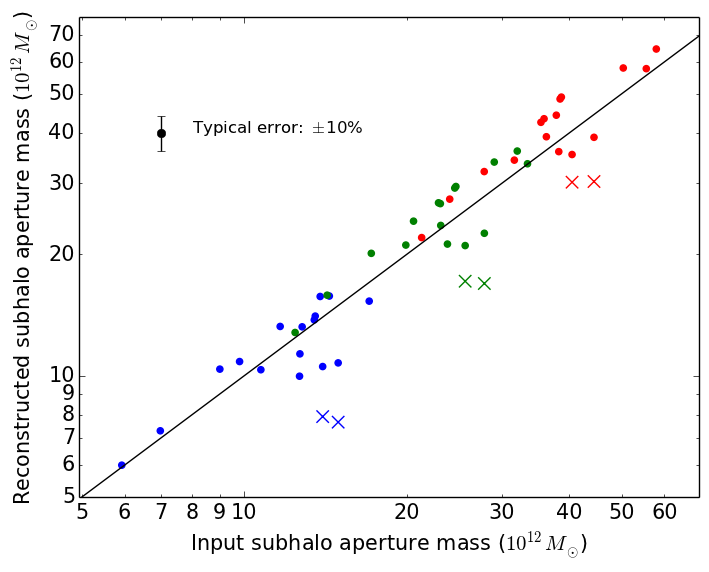} \\
	\caption{The input aperture mass is accurately recovered by the flexion reconstructions.  Color indicate different aperture radii, with blue, green, and red for 10\arcsec, 15\arcsec, and 20\arcsec, respectively.  The line indicates unity, not a fit to the data.  Points marked with an X are measured from reconstructions without flexion, and systematically underestimate the input mass.  The RMS variation of bootstrapped catalog reconstructions indicate a typical aperture mass error of 10\%.}\label{f8}
\end{figure}

In addition to detestability of substructure in terms of identifying individual haloes, we can also quantify the fidelity of our mass reconstructions in terms of how how well the mass maps compare to the input mass distribution by comparing aperture masses.  Figure \ref{f8} plots these results.  We compare aperture masses rather than point-by-point masses since we expect relationship between the input and reconstructed mass distribution to not only include noise propagated from the lensing field estimators, but also an effective convolution kernel that will vary somewhat across the field based on the density of lensing field measurements and the type of measurements in that region (i.e., strong lensing, shear, or flexion).

There is a moderate tightening of the scatter (in dex) as the aperture radius increases, which is consistent with the idea of the reconstruction method imposing and effective convolution kernel just as larger apertures in photometry will omit fewer photons dispersed by a non-trivial point-spread function. Though the measurements for a single halo location with multiple aperture radii are certainly correlated, there is similarly good agreement between the input mass distribution and the reconstructed masses in each of these apertures, confirming that we are in fact reconstructing the substructures on these small scales, as the fraction of the mass enclosed due to the substructure increases as the aperture shrinks.

The measurements scatter about the unity line by 13\% for reconstructions with flexion, which together with the RMS scatter between the different bootstrapped catalog realizations which have an RMS variation of $\sim10\%$ suggest an intrinsic scatter of approximately 8\%.  Without flexion, on the other hand, the reconstructed aperture masses are systematically biased low at the location of the substructures, as we expect given that the substructures are not detected without flexion, by 25-40\%.  This underestimate is also seen in aperture measurements of the main halo masses, showing that the improvement from adding flexion is not only a substructure phenomenon. 

\subsection{Simulation Summary}
We see from these simulations the feasibility for detecting and identifying substructures depends both on their mass and position relative to the cluster halo.  We do not have many detections of substructure at small radii because we select ``un-substructured'' strong lensing image configurations and are therefore artificially enhancing the likelihood that substructure near the Einstein radius of the cluster will remain undetected.  More centrally located subhaloes are also less likely to be detected because of the reduced number of flexion measurements --- again, these substructures are those that would typically be visible in the strong lensing data but do not appear in our data because we select against the configurations that make them apparent.  However, independent of subhalo identification, the masses we measure in apertures as small as 10\arcsec\ are accurate to an error (scatter plus statistical) of 13\% across the reconstructed field.  This is much smaller than the observed systematic offset from not including flexion in the reconstruction.

Flexion data included into the mass reconstructions fill an important information gap in the full lensing data set.  With a mass limit that depends both on the angular size of the subhalo Einstein radius (which is a proxy for the subhalo mass) and its distance from the main cluster halo center, flexion can resolve cluster substructures which are otherwise unobservable from the strong and weak lensing data alone.  Though the specific approach to appropriately identify an individual subhalo in these mass reconstructions has room for optimization, the aperture mass results show that the accuracy of the reconstruction is high, and that reconstructions without flexion systematically underestimate these aperture masses.  This is true even away from the subhalo locations, where flexion information better constrains the overall main halo shape as well.

The three-dimensional halo mass of the substructures at the margins of our detection threshold in this simulation sample have masses of 2-3$\times10^{12}$ M$_\odot$ within an aperture of 10\arcsec, though there is a dependence on the radial position of the subhalo in its detectability.  This demonstrates the efficacy of flexion as a probe of small-scale galaxy cluster substructure.

\section{Discussion}
We have shown that including flexion into the mass reconstruction enhances the sensitivity of the mass reconstruction to substructure a galaxy-cluster-scale lens.  Significantly massive subhaloes which otherwise are undetected in data sets including only multiple image systems and ellipticity measurements are made detectable by including flexion data measured to a precision achievable using current techniques (e.g., CSB).  The additional information from flexion requires no additional observational investment - single-orbit \emph{HST} observations are sufficient for including flexion into the lensing analysis and the mass reconstruction for a typical cluster.  

This result indicates that the addition of flexion into a lensing mass reconstruction better constrains the formation and structure of galaxy clusters and the subhalo mass function.  The work here shows the distinct possibility that dark subhaloes in galaxy clusters of significant mass, such as those included in the simulated measurements we present here, are not detected in lensing mass maps produced without flexion.  Constraints on the subhalo mass function depend on the number density and mass of the detected subhaloes within a cluster, as well as a thorough understanding of the detection limits for subhaloes of a given mass.  By not including flexion, the number density of detected substructures will only be a lower limit on the true substructure density.  With such large structures (up to roughly 10\% the mass of the main halo) it is unlikely, though not impossible, that a cluster would have that much mass in substructure \citep{2010MNRAS.404..502G}. Given that substructures this large were not detectable without flexion, to constrain smaller amounts of substructure with confidence requires that flexion data be included.

The possibility of having significant substructure in a galaxy cluster field  which is otherwise undetected by strong or weak lensing analyses also has important implications on the inferences drawn from the lensing mass distributions.  For example, recent work using galaxy clusters as ``cosmic telescopes'' to select and study high-redshift ($z\gtrsim7$) galaxies \citep[e.g.][]{Hall:2011tk,2012ApJS..199...25P,2012Natur.489..406Z,2012ApJ...747L...9Z} and constraining the properties galaxy population formed at or shortly after the epoch of reionization requires an accurate magnification map of each cluster field.  The overall cluster halo amplifies the effect of substructures on the resulting magnification map, as much of the cluster is typically very near the lensing critical density. By including flexion into the reconstruction and more accurately constraining the lensing potential, the systematic error in the inferred intrinsic properties of the high-redshift galaxy population from unresolved substructures is reduced.

Using Simulation 1 as an example, as compared to an identical main halo without substructure, a \emph{HST/ACS} field containing the substructured halo lens would probe only $\sim88$\% of the solid angle that would be inferred to have been probed if the single halo were the only one reconstructed.  Though the magnitude of this systematic error in each case is dependent on the exact position and mass of the substructures, it is not unreasonable to estimate that for any clusters analyzed without flexion, there is a 10-15\% systematic uncertainty in the solid angle, and therefore the differential volume, probed at any given source redshift.

Furthermore, the intrinsic luminosity determined for any detected high-redshift objects is dependent on the local absolute-value of the magnification determined by the lens reconstruction.  Again referring to the comparison between our simulated lens with substructure and the halo without substructure, the average absolute magnification across a \emph{HST/ACS} field for an infinitely distant source is $\sim50$\% higher for the substructured lens, which would correspond to an average shift of 0.45 magnitude.  Locally the variations between the two lens models is much larger, and in some regions will magnify while in other demagnify a source if the substructures are detected.  

The distribution of local variations in absolute magnification has a long tail corresponding to deviations in the critical curves which would most likely only affect strongly lensed/multiply imaged sources in the $z\gtrsim7$ population.  In the rest of the field, where a large number of the observed high-redshift galaxy sources are likely to be observed, the ratio of the substructured lens magnification to the single-halo lens is well-correlated to the substructured lens magnification ($\rho\approx0.75$).  This means that in regions where the magnification is more enhanced by the substructure, i.e., where magnification maps without flexion will more likely be erroneous, non-detection of substructure and the systematic error in the magnification map produced by that non-detection will create a larger error in the inferred intrinsic luminosity of the object.  Typical magnification ratios in regions near the substructures are $|\mu_{sub}/\mu_{single}|\sim5$, which would create a shift of $\sim1.75$ magnitude in the intrinsic source magnitude inferred without substructure.  That said, there is a significant spread to be expected in these magnification ratios based on the image position relative to the critical curve. The Einstein radius of the substructure imbedded in the main halo in our simulations vary from about 3\arcsec-10\arcsec, meaning that individual detections can be even more than the typical factor and, for lensed image positions near the centers larger substructures the inferred magnification could in fact be lower than expected from a non-substructures lens model.  This type of systematic error would significantly change the resulting luminosity function, as well as the inferred properties of the lensed sources.

Lensing can also determine cluster mass distributions to be compared with other mass estimators.  In particular, the Sunyaev-Zel'dovich decrement and the X-ray surface brightness of a cluster, as functions of position, also constrain the mass distribution of the cluster under the assumption of hydrostatic equilibrium for the intracluster medium.  N-body simulations show that non-thermal pressure support, from cosmic rays and bulk gas motion, plays a significant role in the pressure budget of the cluster, accounting for up to half of the total pressure supporting the cluster \citep{2014ApJ...792...25N}.  While these effects can be accounted for statistically for large cluster samples, doing so on the level of substructures within the cluster requires additional information.  Cluster mass profiles, with the substructure appropriately constrained using flexion, provide an essential test for the hydrostatic assumption, allowing the fractional non-thermal pressure support to be directly quantified for individual clusters.

Flexion is an important addition to the variety of lensing information available from detailed imaging data.  The results presented here strongly motivate the application of this reconstruction technique to simulated data sets with more realistic mass distributions than the simple halo profiles we employ, and additionally applying it to real data.  And while there remains important work to be done to make flexion as robust as the more mature lensing techniques (strong lensing and weak lensing shear), e.g.~ improving our measurement of intrinsic shape scatter in flexion measurements, better understanding the effects of different source selection on the flexion noise properties, etc., we have shown that this work is well worth the undertaking for the resulting enhanced accuracy of reconstructed mass and magnification maps that come from including flexion.  

\section*{Acknowledgements}
Based on observations made with the NASA/ESA Hubble Space Telescope, obtained at the Space Telescope Science Institute, which is operated by the Association of Universities for Research in Astronomy, Inc., under NASA contract NAS 5-26555 and NNX08AD79G. Support for this work was  provided by NASA through
\emph{HST}-AR-13238 and \emph{HST}-AR-13235 from STScI. The authors gratefully acknowledge Paul Schechter for important suggestions and mentorship leading to this work.

% CITATIONS!!!!!!!!!!!!1111!!!!!!1!!1!!1!!1
\bibliographystyle{mn2e}
\bibliography{ms_v8}

\begin{thebibliography}{}
\makeatletter
\relax
\def\mn@urlcharsother{\let\do\@makeother \do\$\do\&\do\#\do\^\do\_\do\%\do\~}
\def\mn@doi{\begingroup\mn@urlcharsother \@ifnextchar [ {\mn@doi@}
  {\mn@doi@[]}}
\def\mn@doi@[#1]#2{\def\@tempa{#1}\ifx\@tempa\@empty \href
  {http://dx.doi.org/#2} {doi:#2}\else \href {http://dx.doi.org/#2} {#1}\fi
  \endgroup}
\def\mn@eprint#1#2{\mn@eprint@#1:#2::\@nil}
\def\mn@eprint@arXiv#1{\href {http://arxiv.org/abs/#1} {{\tt arXiv:#1}}}
\def\mn@eprint@dblp#1{\href {http://dblp.uni-trier.de/rec/bibtex/#1.xml}
  {dblp:#1}}
\def\mn@eprint@#1:#2:#3:#4\@nil{\def\@tempa {#1}\def\@tempb {#2}\def\@tempc
  {#3}\ifx \@tempc \@empty \let \@tempc \@tempb \let \@tempb \@tempa \fi \ifx
  \@tempb \@empty \def\@tempb {arXiv}\fi \@ifundefined
  {mn@eprint@\@tempb}{\@tempb:\@tempc}{\expandafter \expandafter \csname
  mn@eprint@\@tempb\endcsname \expandafter{\@tempc}}}

\bibitem[\protect\citeauthoryear{Bartelmann \& Schneider}{Bartelmann \&
  Schneider}{2001}]{2001PhR...340..291B}
Bartelmann M.,  Schneider P.,  2001, Physics Reports, 340, 291

\bibitem[\protect\citeauthoryear{{Bertin} \& {Arnouts}}{{Bertin} \&
  {Arnouts}}{1996}]{1996A&AS..117..393B}
{Bertin} E.,  {Arnouts} S.,  1996, \aaps, \href
  {http://adsabs.harvard.edu/abs/1996A%26AS..117..393B} {117, 393}

\bibitem[\protect\citeauthoryear{Brada{\v c}, Lombardi  \& Schneider}{Brada{\v
  c} et~al.}{2004}]{2004A&A...424...13B}
Brada{\v c} M.,  Lombardi M.,   Schneider P.,  2004, Astronomy and
  Astrophysics, 424, 13

\bibitem[\protect\citeauthoryear{Brada{\v c} et~al.,}{Brada{\v c}
  et~al.}{2005}]{Bradac:2005ex}
Brada{\v c} M.,  et~al., 2005, Astronomy and Astrophysics, 437, 49

\bibitem[\protect\citeauthoryear{Brada{\v c} et~al.,}{Brada{\v c}
  et~al.}{2006}]{2006ApJ...652..937B}
Brada{\v c} M.,  et~al., 2006, The Astrophysical Journal, 652, 937

\bibitem[\protect\citeauthoryear{Brada{\v c} et~al.,}{Brada{\v c}
  et~al.}{2009}]{Bradac:2009bk}
Brada{\v c} M.,  et~al., 2009, The Astrophysical Journal, 706, 1201

\bibitem[\protect\citeauthoryear{{Brada{\v c}}, {Schneider}, {Lombardi}  \&
  {Erben}}{{Brada{\v c}} et~al.}{2005}]{2005A&A...437...39B}
{Brada{\v c}} M.,  {Schneider} P.,  {Lombardi} M.,   {Erben} T.,  2005, \mn@doi
  [\aap] {10.1051/0004-6361:20042233}, \href
  {http://adsabs.harvard.edu/abs/2005A%26A...437...39B} {437, 39}

\bibitem[\protect\citeauthoryear{Brainerd, Blandford  \& Smail}{Brainerd
  et~al.}{1996}]{1996ApJ...466..623B}
Brainerd T.~G.,  Blandford R.~D.,   Smail I.,  1996, Astrophysical Journal
  v.466, 466, 623

\bibitem[\protect\citeauthoryear{Cain, Schechter  \& Bautz}{Cain
  et~al.}{2011}]{2011ApJ...736...43C}
Cain B.,  Schechter P.~L.,   Bautz M.~W.,  2011, The Astrophysical Journal,
  736, 43

\bibitem[\protect\citeauthoryear{Clowe, Brada{\v c}, Gonzalez, Markevitch,
  Randall, Jones  \& Zaritsky}{Clowe et~al.}{2006}]{2006ApJ...648L.109C}
Clowe D.,  Brada{\v c} M.,  Gonzalez A.~H.,  Markevitch M.,  Randall S.~W.,
  Jones C.,   Zaritsky D.,  2006, The Astrophysical Journal, 648, L109

\bibitem[\protect\citeauthoryear{{Eisenstein} et~al.,}{{Eisenstein}
  et~al.}{2005}]{2005ApJ...633..560E}
{Eisenstein} D.~J.,  et~al., 2005, \mn@doi [\apj] {10.1086/466512}, \href
  {http://adsabs.harvard.edu/abs/2005ApJ...633..560E} {633, 560}

\bibitem[\protect\citeauthoryear{Er, Li  \& Schneider}{Er
  et~al.}{2010}]{Er:2010uo}
Er X.,  Li G.,   Schneider P.,  2010, arXiv.org

\bibitem[\protect\citeauthoryear{Giocoli, Tormen, Sheth  \& van~den
  Bosch}{Giocoli et~al.}{2010}]{2010MNRAS.404..502G}
Giocoli C.,  Tormen G.,  Sheth R.~K.,   van~den Bosch F.~C.,  2010, Monthly
  Notices of the Royal Astronomical Society, 404, 502

\bibitem[\protect\citeauthoryear{{Giodini}, {Lovisari}, {Pointecouteau},
  {Ettori}, {Reiprich}  \& {Hoekstra}}{{Giodini}
  et~al.}{2013}]{2013SSRv..177..247G}
{Giodini} S.,  {Lovisari} L.,  {Pointecouteau} E.,  {Ettori} S.,  {Reiprich}
  T.~H.,   {Hoekstra} H.,  2013, \mn@doi [\ssr] {10.1007/s11214-013-9994-5},
  \href {http://adsabs.harvard.edu/abs/2013SSRv..177..247G} {177, 247}

\bibitem[\protect\citeauthoryear{Goldberg \& Bacon}{Goldberg \&
  Bacon}{2005}]{2005ApJ...619..741G}
Goldberg D.~M.,  Bacon D.~J.,  2005, The Astrophysical Journal, 619, 741

\bibitem[\protect\citeauthoryear{Goldberg \& Leonard}{Goldberg \&
  Leonard}{2007}]{2007ApJ...660.1003G}
Goldberg D.~M.,  Leonard A.,  2007, The Astrophysical Journal, 660, 1003

\bibitem[\protect\citeauthoryear{Hall et~al.,}{Hall et~al.}{2011}]{Hall:2011tk}
Hall N.,  et~al., 2011, arXiv.org

\bibitem[\protect\citeauthoryear{Hinshaw et~al.,}{Hinshaw
  et~al.}{2013}]{2013ApJS..208...19H}
Hinshaw G.,  et~al., 2013, The Astrophysical Journal Supplement, 208, 19

\bibitem[\protect\citeauthoryear{{Kowalski} et~al.,}{{Kowalski}
  et~al.}{2008}]{2008ApJ...686..749K}
{Kowalski} M.,  et~al., 2008, \mn@doi [\apj] {10.1086/589937}, \href
  {http://adsabs.harvard.edu/abs/2008ApJ...686..749K} {686, 749}

\bibitem[\protect\citeauthoryear{{Kravtsov}, {Berlind}, {Wechsler}, {Klypin},
  {Gottl{\"o}ber}, {Allgood}  \& {Primack}}{{Kravtsov}
  et~al.}{2004}]{2004ApJ...609...35K}
{Kravtsov} A.~V.,  {Berlind} A.~A.,  {Wechsler} R.~H.,  {Klypin} A.~A.,
  {Gottl{\"o}ber} S.,  {Allgood} B.,   {Primack} J.~R.,  2004, \mn@doi [\apj]
  {10.1086/420959}, \href {http://adsabs.harvard.edu/abs/2004ApJ...609...35K}
  {609, 35}

\bibitem[\protect\citeauthoryear{{Melchior}, {B{\"o}hnert}, {Lombardi}  \&
  {Bartelmann}}{{Melchior} et~al.}{2010}]{2010A&A...510A..75M}
{Melchior} P.,  {B{\"o}hnert} A.,  {Lombardi} M.,   {Bartelmann} M.,  2010,
  \mn@doi [\aap] {10.1051/0004-6361/200912785}, \href
  {http://adsabs.harvard.edu/abs/2010A%26A...510A..75M} {510, A75}

\bibitem[\protect\citeauthoryear{{Moore}, {Ghigna}, {Governato}, {Lake},
  {Quinn}, {Stadel}  \& {Tozzi}}{{Moore} et~al.}{1999}]{1999ApJ...524L..19M}
{Moore} B.,  {Ghigna} S.,  {Governato} F.,  {Lake} G.,  {Quinn} T.,  {Stadel}
  J.,   {Tozzi} P.,  1999, \mn@doi [\apjl] {10.1086/312287}, \href
  {http://adsabs.harvard.edu/abs/1999ApJ...524L..19M} {524, L19}

\bibitem[\protect\citeauthoryear{{Nagai} \& {Kravtsov}}{{Nagai} \&
  {Kravtsov}}{2005}]{2005ApJ...618..557N}
{Nagai} D.,  {Kravtsov} A.~V.,  2005, \mn@doi [\apj] {10.1086/426016}, \href
  {http://adsabs.harvard.edu/abs/2005ApJ...618..557N} {618, 557}

\bibitem[\protect\citeauthoryear{{Natarajan}, {De Lucia}  \&
  {Springel}}{{Natarajan} et~al.}{2007}]{2007MNRAS.376..180N}
{Natarajan} P.,  {De Lucia} G.,   {Springel} V.,  2007, \mn@doi [\mnras]
  {10.1111/j.1365-2966.2007.11399.x}, \href
  {http://adsabs.harvard.edu/abs/2007MNRAS.376..180N} {376, 180}

\bibitem[\protect\citeauthoryear{{Natarajan}, {Kneib}, {Smail}, {Treu},
  {Ellis}, {Moran}, {Limousin}  \& {Czoske}}{{Natarajan}
  et~al.}{2009}]{2009ApJ...693..970N}
{Natarajan} P.,  {Kneib} J.-P.,  {Smail} I.,  {Treu} T.,  {Ellis} R.,  {Moran}
  S.,  {Limousin} M.,   {Czoske} O.,  2009, \mn@doi [\apj]
  {10.1088/0004-637X/693/1/970}, \href
  {http://adsabs.harvard.edu/abs/2009ApJ...693..970N} {693, 970}

\bibitem[\protect\citeauthoryear{{Nelson}, {Lau}  \& {Nagai}}{{Nelson}
  et~al.}{2014}]{2014ApJ...792...25N}
{Nelson} K.,  {Lau} E.~T.,   {Nagai} D.,  2014, \mn@doi [\apj]
  {10.1088/0004-637X/792/1/25}, \href
  {http://adsabs.harvard.edu/abs/2014ApJ...792...25N} {792, 25}

\bibitem[\protect\citeauthoryear{Okura \& Futamase}{Okura \&
  Futamase}{2009}]{Okura:2009cw}
Okura Y.,  Futamase T.,  2009, The Astrophysical Journal, 699, 143

\bibitem[\protect\citeauthoryear{Peter, Rocha, Bullock  \& Kaplinghat}{Peter
  et~al.}{2013}]{2013MNRAS.430..105P}
Peter A. H.~G.,  Rocha M.,  Bullock J.~S.,   Kaplinghat M.,  2013, Monthly
  Notices of the Royal Astronomical Society, 430, 105

\bibitem[\protect\citeauthoryear{{Planck Collaboration} et~al.,}{{Planck
  Collaboration} et~al.}{2013}]{2013arXiv1303.5076P}
{Planck Collaboration} et~al., 2013, arXiv.org, p.~5076

\bibitem[\protect\citeauthoryear{Postman et~al.,}{Postman
  et~al.}{2012}]{2012ApJS..199...25P}
Postman M.,  et~al., 2012, The Astrophysical Journal Supplement, 199, 25

\bibitem[\protect\citeauthoryear{Rocha, Peter, Bullock, Kaplinghat,
  Garrison-Kimmel, O{\~n}orbe  \& Moustakas}{Rocha
  et~al.}{2013}]{2013MNRAS.430...81R}
Rocha M.,  Peter A. H.~G.,  Bullock J.~S.,  Kaplinghat M.,  Garrison-Kimmel S.,
   O{\~n}orbe J.,   Moustakas L.~A.,  2013, Monthly Notices of the Royal
  Astronomical Society, 430, 81

\bibitem[\protect\citeauthoryear{Schneider \& Er}{Schneider \&
  Er}{2008}]{Schneider:2008ho}
Schneider P.,  Er X.,  2008, Astronomy and Astrophysics, 485, 363

\bibitem[\protect\citeauthoryear{Schrabback et~al.,}{Schrabback
  et~al.}{2007}]{2007A&A...468..823S}
Schrabback T.,  et~al., 2007, Astronomy and Astrophysics, 468, 823

\bibitem[\protect\citeauthoryear{{Shaw}, {Weller}, {Ostriker}  \&
  {Bode}}{{Shaw} et~al.}{2006}]{2006ApJ...646..815S}
{Shaw} L.~D.,  {Weller} J.,  {Ostriker} J.~P.,   {Bode} P.,  2006, \mn@doi
  [\apj] {10.1086/505016}, \href
  {http://adsabs.harvard.edu/abs/2006ApJ...646..815S} {646, 815}

\bibitem[\protect\citeauthoryear{Tinker, Kravtsov, Klypin, Abazajian, Warren,
  Yepes, Gottl{\"o}ber  \& Holz}{Tinker et~al.}{2008}]{2008ApJ...688..709T}
Tinker J.,  Kravtsov A.~V.,  Klypin A.,  Abazajian K.,  Warren M.,  Yepes G.,
  Gottl{\"o}ber S.,   Holz D.~E.,  2008, The Astrophysical Journal, 688, 709

\bibitem[\protect\citeauthoryear{{Vegetti}, {Lagattuta}, {McKean}, {Auger},
  {Fassnacht}  \& {Koopmans}}{{Vegetti} et~al.}{2012}]{2012Natur.481..341V}
{Vegetti} S.,  {Lagattuta} D.~J.,  {McKean} J.~P.,  {Auger} M.~W.,  {Fassnacht}
  C.~D.,   {Koopmans} L.~V.~E.,  2012, \mn@doi [\nat] {10.1038/nature10669},
  \href {http://adsabs.harvard.edu/abs/2012Natur.481..341V} {481, 341}

\bibitem[\protect\citeauthoryear{{Zheng} et~al.,}{{Zheng}
  et~al.}{2012}]{2012Natur.489..406Z}
{Zheng} W.,  et~al., 2012, \mn@doi [\nat] {10.1038/nature11446}, \href
  {http://adsabs.harvard.edu/abs/2012Natur.489..406Z} {489, 406}

\bibitem[\protect\citeauthoryear{{Zitrin} et~al.,}{{Zitrin}
  et~al.}{2012}]{2012ApJ...747L...9Z}
{Zitrin} A.,  et~al., 2012, \mn@doi [\apjl] {10.1088/2041-8205/747/1/L9}, \href
  {http://adsabs.harvard.edu/abs/2012ApJ...747L...9Z} {747, L9}

\makeatother
\end{thebibliography}

\appendix
\section{Extended Aperture Mass Results}
Table \ref{apmasstable} shows input and measured masses for each subhalo, along with the distance from each subhalo to the cluster halo center.  We include a three-dimensional mass for the subhalo alone, as well as the mass within the aperture for the full input mass distribution, and the reconstructed mass distribution.  Note that the aperture masses include contribution from both the subhalo and the main halo.  Values are tabulated at different aperture radii - 10\arcsec, 15\arcsec, and 20\arcsec.  We also include aperture mass measurements for the locations of the subhaloes in Simulation 1 as reconstructed without flexion.

\begin{sidewaystable}
\begin{minipage}{\linewidth}\vspace{3in}
\caption{Masses for each simulation subhalo at three different radii: the three dimensional input mass for the subhalo alone, along with input and measured aperture masses. * denotes reconstructions without flexion)}\label{apmasstable}
\begin{tabular}{lccccccccccc}
\hline\hline
SIM/HALO & Detection? & Dist.~to main halo & $M_{3D,10\arcsec}$ & $M_{10\arcsec,input}$ & $M_{10\arcsec,meas}$ & $M_{3D,15\arcsec}$ & $M_{15\arcsec,input}$ & $M_{15\arcsec,meas}$ & $M_{3D,20\arcsec}$ & $M_{20\arcsec,input}$ & $M_{20\arcsec,meas}$ \\
 & & (arcsec) & ($10^{12} M_\odot$) & ($10^{12} M_\odot$) & ($10^{12} M_\odot$) & ($10^{12} M_\odot$) & ($10^{12} M_\odot$) & ($10^{12} M_\odot$) & ($10^{12} M_\odot$) & ($10^{12} M_\odot$) & ($10^{12} M_\odot$) \\
\hline
Sim.~1 - A$^*$	& N 		& 55.5 & 7.92 & 14.92 &  7.67 & 12.00 & 27.84 & 16.93 & 16.09 & 44.44 & 30.31 \\ 
Sim.~1 - A		& Y 		& 55.5 & 7.92 & 14.92 & 10.75 & 12.00 & 27.84 & 22.55 & 16.09 & 44.44 & 39.01 \\
Sim.~1 - B$^*$	& N 		& 64.9 & 7.92 & 13.97 &  7.91 & 12.00 & 25.65 & 17.15 & 16.09 & 40.47 & 30.19 \\ 
Sim.~1 - B		& Y 		& 64.9 & 7.92 & 13.97 & 10.53 & 12.00 & 25.65 & 21.02 & 16.09 & 40.47 & 35.36 \\ 
Sim.~2 - A 		& Y 		& 70.9 & 7.92 & 13.47 & 13.75 & 12.00 & 24.54 & 29.20 & 16.09 & 38.45 & 48.56 \\ 
Sim.~2 - B 		& Y 		& 70.2 & 7.92 & 13.53 & 14.05 & 12.00 & 24.65 & 29.45 & 16.09 & 38.67 & 49.08 \\ 
Sim.~3 - A 		& N 		& 29.0 & 1.29 & 14.37 & 15.75 &  1.96 & 32.03 & 36.07 &  2.62 & 57.97 & 64.60 \\ 
Sim.~3 - B 		& N 		& 76.7 & 2.06 &  6.99 &  7.30 &  3.12 & 14.24 & 15.83 &  4.18 & 24.01 & 27.40 \\ 
Sim.~4 - A 		& marginal 	& 50.6 & 2.14 &  9.80 & 10.84 &  3.24 & 20.58 & 24.16 &  4.35 & 35.44 & 42.50 \\ 
Sim.~4 - B 		& Y 		& 71.5 & 7.39 & 12.67 & 11.32 & 11.20 & 23.11 & 23.58 & 15.01 & 36.27 & 39.14 \\ 
Sim.~5 - A 		& marginal 	& 35.7 & 3.23 & 13.82 & 15.72 &  4.89 & 29.04 & 33.87 &  6.55 & 50.35 & 57.96 \\ 
Sim.~5 - B 		& Y 		& 79.6 & 4.22 &  9.01 & 10.38 &  6.40 & 17.19 & 20.11 &  8.57 & 27.82 & 32.08 \\ 
Sim.~6 - A 		& Y 		& 76.6 & 7.80 & 12.80 & 13.21 & 11.81 & 23.07 & 26.71 & 15.83 & 35.91 & 43.38 \\ 
Sim.~6 - B 		& Y 		& 55.3 & 4.63 & 11.65 & 13.24 &  7.01 & 22.89 & 26.83 &  9.39 & 37.84 & 44.26 \\ 
\hline
\end{tabular}
\end{minipage}
\end{sidewaystable}

\end{document}